\documentclass[twoside,twocolumn,9pt]{article}
\usepackage{extsizes}
\usepackage[super,sort&compress,comma]{natbib} 
\usepackage[version=3]{mhchem}
\usepackage[left=1.5cm, right=1.5cm, top=1.785cm, bottom=2.0cm]{geometry}
\usepackage{balance}
\usepackage{times,mathptmx}
\usepackage{sectsty}
\usepackage{graphicx} 
\usepackage{lastpage}
\usepackage[format=plain,justification=justified,singlelinecheck=false,font={stretch=1.125,small,sf},labelfont=bf,labelsep=space]{caption}
\usepackage{float}
\usepackage{fancyhdr}
\usepackage{amssymb}
\usepackage{fnpos}
\usepackage[english]{babel}
\addto{\captionsenglish}{%
  
}
\usepackage{array}
\usepackage{droidsans}
\usepackage{charter}
\usepackage[T1]{fontenc}
\usepackage[usenames,dvipsnames]{xcolor}
\usepackage{setspace}
\usepackage[compact]{titlesec}
\usepackage{hyperref}

\usepackage{epstopdf}

\usepackage{textcomp}

\newcommand{\p}{\partial}
\newcommand{\ep}{\epsilon}
\newcommand{\bkap}{\vec{\kappa}}

\definecolor{cream}{RGB}{222,217,201}

\begin{document}

\pagestyle{fancy}
\thispagestyle{plain}
\fancypagestyle{plain}{

\renewcommand{\headrulewidth}{0pt}
}

\makeFNbottom
\makeatletter
\renewcommand\LARGE{\@setfontsize\LARGE{15pt}{17}}
\renewcommand\Large{\@setfontsize\Large{12pt}{14}}
\renewcommand\large{\@setfontsize\large{10pt}{12}}
\renewcommand\footnotesize{\@setfontsize\footnotesize{7pt}{10}}
\makeatother

\renewcommand{\thefootnote}{\fnsymbol{footnote}}
\renewcommand\footnoterule{\vspace*{1pt}%
\color{cream}\hrule width 3.5in height 0.4pt \color{black}\vspace*{5pt}} 
\setcounter{secnumdepth}{5}

\makeatletter 
\renewcommand\@biblabel[1]{#1}            
\renewcommand\@makefntext[1]%
{\noindent\makebox[0pt][r]{\@thefnmark\,}#1}
\makeatother 
\renewcommand{\figurename}{\small{Fig.}~}
\sectionfont{\sffamily\Large}
\subsectionfont{\normalsize}
\subsubsectionfont{\bf}
\setstretch{1.125} 
\setlength{\skip\footins}{0.8cm}
\setlength{\footnotesep}{0.25cm}
\setlength{\jot}{10pt}
\titlespacing*{\section}{0pt}{4pt}{4pt}
\titlespacing*{\subsection}{0pt}{15pt}{1pt}

\fancyfoot{}
\fancyfoot[LO,RE]{\vspace{-7.1pt}\includegraphics[height=9pt]{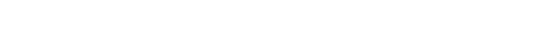}}
\fancyfoot[CO]{\vspace{-7.1pt}\hspace{13.2cm}\includegraphics{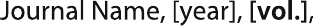}}
\fancyfoot[CE]{\vspace{-7.2pt}\hspace{-14.2cm}\includegraphics{head_foot/RF}}
\fancyfoot[RO]{\footnotesize{\sffamily{1--\pageref{LastPage} ~\textbar  \hspace{2pt}\thepage}}}
\fancyfoot[LE]{\footnotesize{\sffamily{\thepage~\textbar\hspace{3.45cm} 1--\pageref{LastPage}}}}
\fancyhead{}
\renewcommand{\headrulewidth}{0pt} 
\renewcommand{\footrulewidth}{0pt}
\setlength{\arrayrulewidth}{1pt}
\setlength{\columnsep}{6.5mm}
\setlength\bibsep{1pt}

\makeatletter 
\newlength{\figrulesep} 
\setlength{\figrulesep}{0.5\textfloatsep} 

\newcommand{\topfigrule}{\vspace*{-1pt}%
\noindent{\color{cream}\rule[-\figrulesep]{\columnwidth}{1.5pt}} }

\newcommand{\botfigrule}{\vspace*{-2pt}%
\noindent{\color{cream}\rule[\figrulesep]{\columnwidth}{1.5pt}} }

\newcommand{\dblfigrule}{\vspace*{-1pt}%
\noindent{\color{cream}\rule[-\figrulesep]{\textwidth}{1.5pt}} }

\makeatother

\twocolumn[
  \begin{@twocolumnfalse}
\vspace{3cm}
\sffamily
\begin{tabular}{m{4.5cm} p{13.5cm} }

\includegraphics{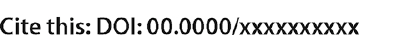} & \noindent\LARGE{\textbf{Dynamics of Fiberboids}}\\ 
\vspace{0.3cm} & \vspace{0.3cm} \\

 & \noindent\large{Antony Bazir,
 \textit{$^{a}$} 
 Arthur Baumann,\textit{$^{b}$} 
 Falko Ziebert,\textit{$^{c}$} and 
 Igor M. Kuli\'{c} \textit{$^{b,d}$}} \\

\includegraphics{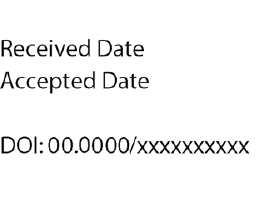} & \noindent\normalsize{Fiberboids 
are active filaments trapped at the interface of two phases,
able of harnessing energy (and matter) fluxes across the interface
in order to produce a rolling-like self-propulsion. We discuss
several table-top examples and develop the physical framework
for understanding their complex dynamics. In spite of some specific
features in the examples studied we conclude that the phenomenon of
fiberboids is highly generic and robust across different materials, types of fluxes and timescales. 
Fiberboid motility should play a role from the macroscopic realm
down to the micro scale and, as recently hypothesized, possibly as a means of biological self-propulsion that has escaped previous attention.
} \\

\end{tabular}

 \end{@twocolumnfalse} \vspace{0.6cm}

  ]

\renewcommand*\rmdefault{bch}\normalfont\upshape
\rmfamily
\vspace{-1cm}


\footnotetext{\textit{$^{a}$~Institut Lumi\`ere Mati\`ere, 
UMR 5306, Universit\'e Lyon 1-CNRS, 69622 Villeurbanne, France. }}
\footnotetext{\textit{$^{b}$~Institut Charles Sadron, CNRS, 23 rue du Loess, BP 84047,
67034 Strasbourg, France.
}}
\footnotetext{\textit{$^{c}$~Institute for Theoretical Physics, Heidelberg University, 
Philosophenweg 19, 69120 Heidelberg, Germany; 
E-mail: ziebert@thphys.uni-heidelberg.de}}
\footnotetext{\textit{$^{d}$~Leibniz Institute for Polymer Research (IPF), 01069 Dresden,
Germany; E-mail: kulic@unistra.fr}}

\section{Introduction}
Self-propulsion, the motion of objects using internal or external energy sources or fluxes,
is a hallmark of living objects and an active subject of nonequilibrium 
physics and soft matter research. While bacterial motion \cite{microswimmers} inspired artificial microswimmers 
have been abundantly proposed and investigated \cite{ActiveParticles}, 
examples for man-made substrate-based self-propellers are still rather scarce.
Examples on the micro-scale comprise Quincke \cite{QuinckeLyon} and magnetic (fluid interface) rollers \cite{magn_rollers}
and on the macro-scale shaken granular rods \cite{Kudrolli},
Marangoni-driven droplets \cite{Dauchotdroplets} or
Leidenfrost wheels \cite{Querewheel}.
We here would like to add a whole class of more examples to the latter, 
using a very simple and generic strategy to motorize fibers
confined at non-equilibrium interfaces \cite{Baumann}.

\begin{figure*}
\centering
\includegraphics[width=17cm]{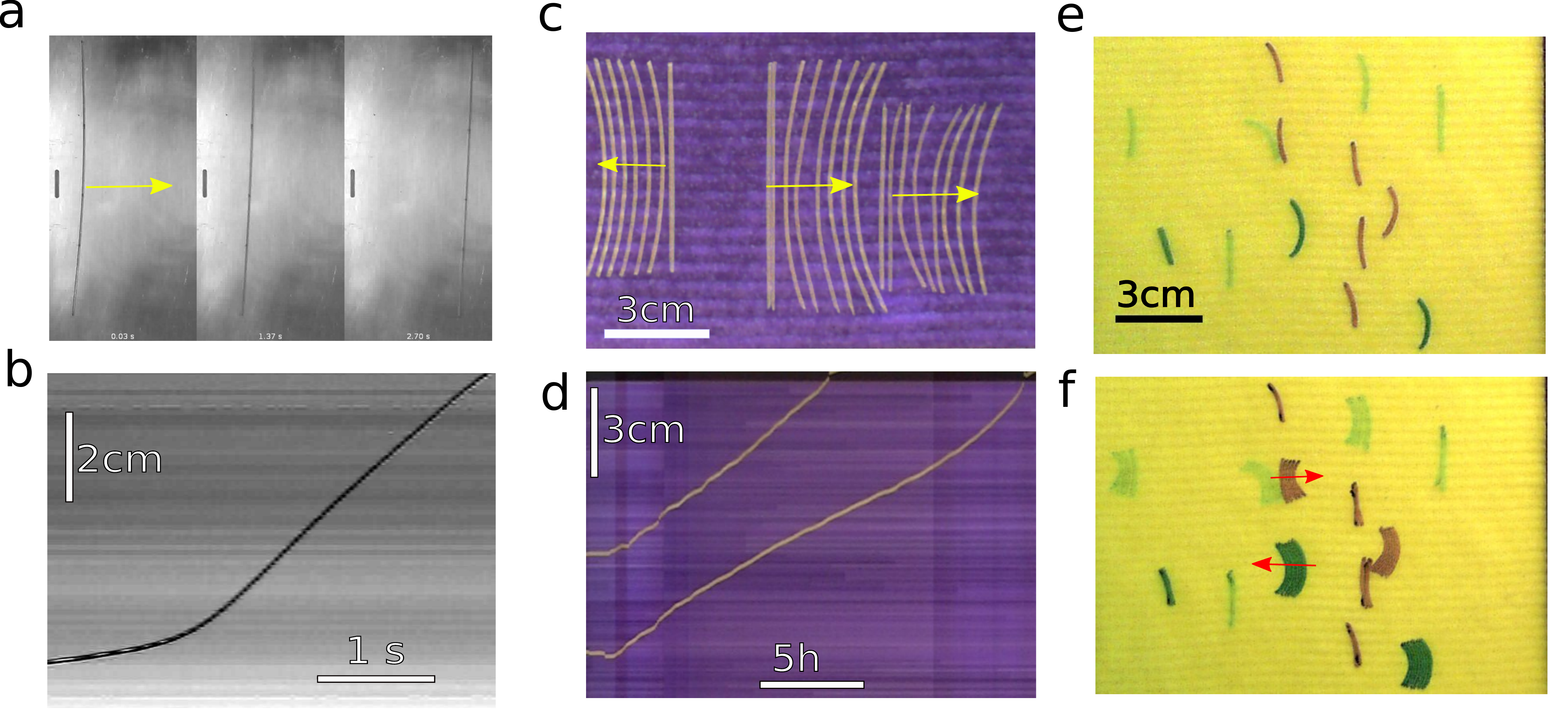}
\caption{\label{fig_Fiberboids-in-action}
Fiberboids in action. 
a) A nylon fiber rolling on a heated aluminum substrate. See supplementary movie 1.
b) A kymograph (space vs.~time plot) of the center of mass position of the nylon fiber
shown in a). 
c) A stroboscopic picture of three initially dry spaghetti
curving and then rolling on a wet towel (at $\sim60$\% air humidity at $25^\circ$C). See supplementary movie 2.
d) Kymographs for two of the rolling spaghetti shown in c). 
e) Polyacrylamide (PAM) hydrogel rods on a wet towel. 
f) Stroboscopic picture showing the rolling motion in e). 
}
\end{figure*}

When two bulk phases meet at an interface, more often than not, they
are in a mutual thermodynamic disequilibrium. As a consequence, they
exchange fluxes of energy and matter along the normal to the interface.
If now a deformable filament is trapped in the interface plane, and
if it is elastically responsive to the agent flowing across it, something
truly astonishing happens: the filament bends into an arc, starts
to axially ``spin'' and move along the interface, transforming itself
from a passive object into a simple engine, which we here term ``fiberboid''.
Some extremely simple, table-top experimental realizations of fiberboids are
shown in Fig.~\ref{fig_Fiberboids-in-action}.

The bidirectional active motion of fiberboids  (albeit with a focus on the unidirectional rotational motion of such fibers closed into a torus)
was described in a recent work\cite{Baumann}. 
Yet our understanding of their general physics, in particular their dynamics, 
remains incomplete. Due to their self-organized nature that involves 
symmetry breaking, self-propulsion, bi-directionality of motion, 
interactions and collective effects,
fiberboids are significantly more complex than their unidirectional
variants closed into a torus or spirals \cite{Baumann}.
Yet, at the same time, fiberboids are extremely simple to generate
as demonstrated by the examples shown in Fig.~\ref{fig_Fiberboids-in-action} 
and detailed in the next section.

In this work
we will distill the generic features of fiberboid motility and
contrast them to more particular features exhibited by specific experimental
realizations. The outline of this work is as follows. 
After discussing two distinct examples in section \ref{examples}, 
in section \ref{generic_model} we develop a simple dynamical model describing fiberboid motion. 
Comparing the predictions of this toy model to experimental data, 
we will see how specific observations for different fiberboid types naturally lead to 
model extensions. In section \ref{social} we study the ``social''
behavior of fiberboids, i.e.~how they interact
with obstacles and other fiberboids, paving the way to studying their
collective behavior in the near future.

\section{Examples}\label{examples}
To get a taste for the fiberboid phenomenon, let us start with two easy
to reproduce table-top examples. Fig.~\ref{fig_Fiberboids-in-action}a)
shows snapshots of a nylon fiber -- a piece of commercial fishing line --
on an aluminum plate heated at $\sim140^\circ$C. Fig.~\ref{fig_Fiberboids-in-action}b)
displays a so-called kymograph (space-time plot generated from time-lapse
images as shown in part a), clearly evidencing
a rolling motion along the substrate with a typical velocity of few cm/s. 
The direction of motion with respect to the fiber curvature 
depends on the coupling of the thermal gradient through the fiber 
(between the heated plate and the colder surroundings)
to the fibers elastic deformation, cf.~Fig.~\ref{fig_Sketch},
i.e.~it is determined by the sign of the coefficient of thermal expansion.  
For thermally contracting materials as nylon, the fiber curves away from the direction of motion,
while for thermally expanding materials,
such as polydimethylsiloxane (PDMS) fibers, it curves into the direction of motion\cite{Baumann}.
The same effect has been recently applied to propel mono-domain liquid
crystalline elastomer rods \cite{Cai}. 

Figs.~\ref{fig_Fiberboids-in-action}c)-f) show a second possibility
to drive fibers: here, uncooked spaghetti (c), as well as Polyacrylamide (PAM)
hydrogel rods (e), were put on a wet kitchen towel. 
The water then slowly enters the rods via the wet substrate, differentially swells 
and hence curves them. As in the case of heat flow in the first example, 
there is a constant solvent flow (water enters from the bottom and evaporates at the top)
and the fibers start to roll, albeit much slower here, with typical velocities 
of few cm/hour, cf.~the kymograph in Fig.~\ref{fig_Fiberboids-in-action}d). 
The traveling direction with respect to curvature for hygroscopic swelling 
is analogous to the thermal expansion case. A related effect was reported earlier
using PDMS rods \cite{Ghatak,Ghatak2}. However, there, a solvent
droplet was injected at what later becomes the back of the rolling rod,
i.e.~the spontaneous symmetry breaking and bidirectionality,
characteristic for fiberboids (cf.~Ref.~\cite{Baumann} and as discussed below), 
was absent in these earlier examples.

\section{Generic models}\label{generic_model}
\begin{figure}
\includegraphics[width=8.2cm]{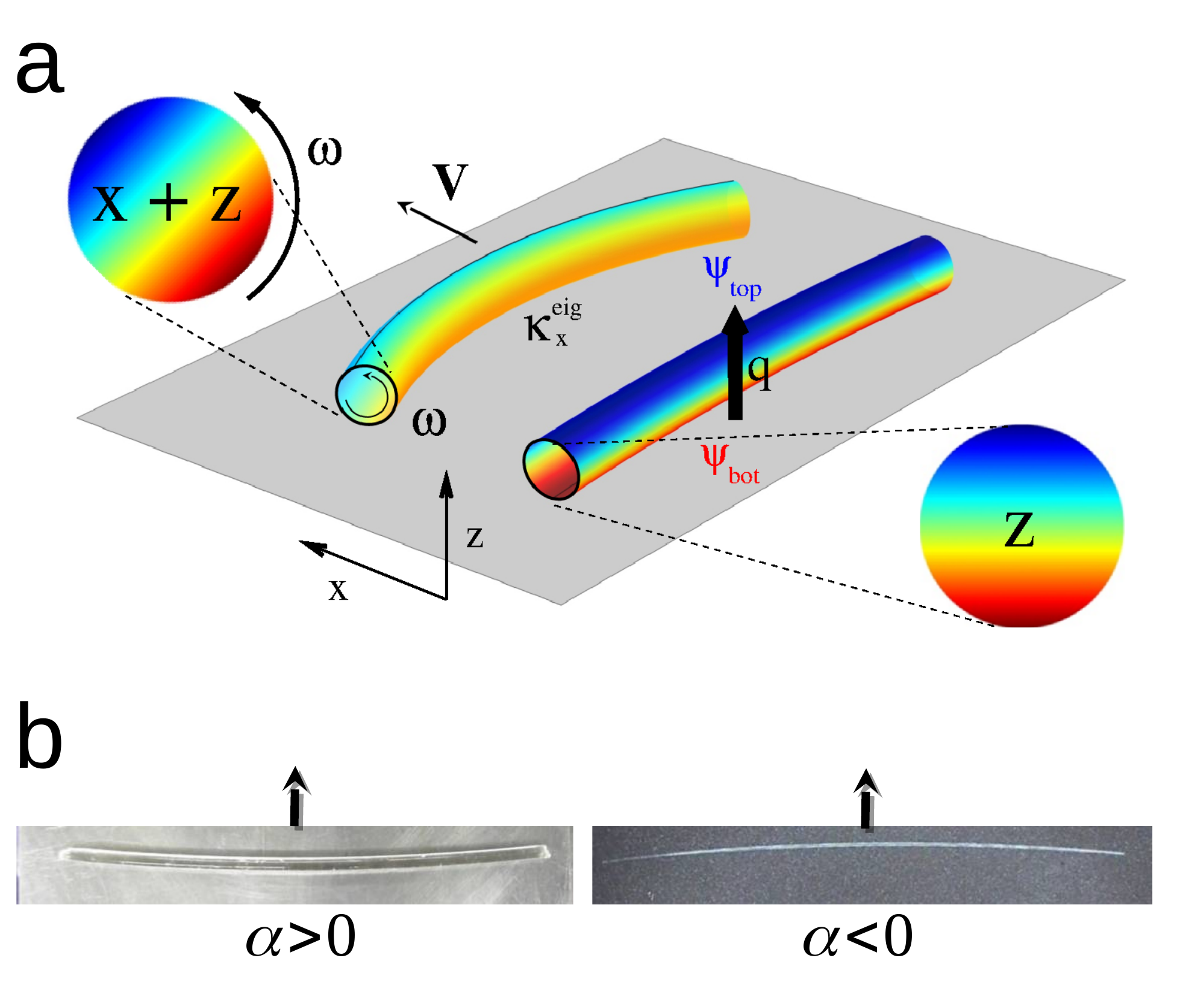}
\caption{ \label{fig_Sketch}
a) Sketch of the mechanism underlying fiberboid rolling. 
The difference in propellant density $\psi_{top}-\psi_{bottom}$
gives rise to a propellant flux establishing a gradient along the z-direction. 
Differential expansion then leads to a preferred out-of-plane eigen-curvature, 
which however cannot be realized due to the confinement to the interface. 
Instead, an in-plane curvature $\kappa_{x}^{eig}$ is built-up and 
the constant phase difference between the propellant-induced
strain and the realized strain results in a driving torque 
that induces a rotation of every cross-section with frequency $\omega$.
b) Depending on their sign of the expansion coefficient $\alpha$,
fiberboids fall into two classes: ``vexers'' having $\alpha>0$
and rolling towards their center of curvature, as exemplified by a PDMS rod. 
And ``cavers'' having $\alpha<0$ and moving away from it, as exemplified for a nylon fiber. 
Arrows show direction of motion.
}
\end{figure}

Looking at these examples, with different driving mechanisms and operating
at (vastly) different timescales, a natural question arises: Is
there a unified approach to understand the dynamics? Both systems
have in common that there is a propelling agent ``flowing'' between
two phases (baths) through the filament. The propellant, which stands
for heat energy density or water concentration, respectively, in the
two example cases, has a scalar density within the material denoted
by $\psi\left(\rho,\phi,t\right)$. The latter depends on the polar
coordinate position $\left(\rho,\phi\right)$ in the fiber cross-section
in the laboratory frame and on time $t$. For the basic physics, however,
only the differences of $\psi$ in the normal (z-direction/top-to-bottom)
and in-plane direction (x-direction/left-to-right) will be important,
cf.~Fig.~\ref{fig_Sketch}. 

The responsiveness of the filament is related to the coupling between
the propellant density and elastic deformations. In the simplest case,
the presence of the propellant gives rise to an isotropic material
eigenstrain (also called prestrain) tensor 
\begin{equation}\label{eigenstrain}
\varepsilon_{ij}^{eig}=\alpha\psi\delta_{ij}, 
\end{equation}
with $i,j\in x,y,z$ and  $\delta_{ij}$ the Kronecker delta. 
Here $\alpha$ is an isotropic linear expansion coefficient, 
describing thermal expansion or gel swelling, respectively. 
Note that this coefficient can also be negative,
corresponding to contraction, as in case of heated nylon. 
The eigenstrain can be seen \cite{Efrati,Eigenstrain} as 
the locally preferred strain in the absence of local
and global constraints. It evolves in time with the propellant concentration
and gives rise to all mechanical effects in fiberboids.

Overall the propellant's evolution is governed by three processes: (i) the
propellant influx rate $q$, (ii) its  relaxation rate $r$
and (iii) the rotation of the cross-section with angular velocity
$\omega$, in case the rolling motion has set in.
In the absence of drive, the propellant evolves via a diffusive process,
heat diffusion and swelling agent diffusion in the two examples. 
By scaling, the associated time scale $\tau$ is determined by the diffusion 
coefficient of the propellant, $D_\psi$,
and the dimension (i.e.~radius) of the fiber cross-section, $R$, defining 
the propellant's diffusive relaxation rate 
\begin{equation}
r=\frac{1}{\tau}\sim D_{\psi}R^{-2}. 
\end{equation}
In the driven case, the influx rate then scales as
\begin{equation}
q\sim D_{\psi}R^{-2}\Delta_{\updownarrow}\psi,
\end{equation}
i.e.~it is given by the same characteristic time scale but in addition is proportional to 
the difference of the propellant 
concentration $\Delta_{\updownarrow}\psi=\psi_{top}-\psi_{bottom}$
(between the top and the bottom of the fiber). 
In the absence of rotation, the propellant diffusive relaxation rate 
$r$ counterbalances this influx, giving rise to a typical z-gradient $\nabla_{z}\psi\sim\frac{q}{Rr}$.
The latter induces an eigenstrain gradient in the axial component
(i.e. $\varepsilon_{yy}^{eig}$, where $y$ lies along the fiber axis),
which in turn leads to a preferred out-of-plane eigen-curvature,
\begin{equation}
\kappa_{z}^{eig}\sim R^{-1}\Delta_{\updownarrow}\varepsilon_{yy}^{eig}.
\end{equation}

This curvature $\kappa_{z}^{eig}$, however, is entirely ``virtual'',
i.e.~it cannot be actually 
realized due to the presence of 
normal forces\footnote{Two generic cases of confinement may occur: In the case of short range,
adhesive contact forces -- like e.g.~Pickering-type pinning at the
interface \cite{pickering} -- the fiber stays fully confined as long as its
bending energy density is smaller than the gain in adhesion energy.
For macroscopic fibers 
confinement is due to gravity, which is a
small and long-ranged force and additional effects come into play that 
can be estimated as follows:
a fiber of length $l$ that is confined to the plane reduces its potential
energy $e_{pot}\sim\rho_0 gl^{3}R^{2}\kappa_{z}^{eig}$, with $\rho_0$
the density and $g$ the gravitational acceleration, but increases
its bending energy by $e_{bend}\sim l\,YR^{4}(\kappa_{z}^{eig})^2$.
Matching the two energies yields a characteristic confinement length
\begin{equation}
l_{conf}=R\left(\frac{Y\kappa_{z}^{eig}}{\rho_0 g}\right)^{1/2}.\nonumber
\end{equation}
Hence, unlike for the contact force case, where the whole fiber is
either fully confined or not, in the gravity case both ends of the
fiber are always unconfined on a typical length scale $l_{conf}$.
For long fibers with $l\gg l_{conf}$ these boundary effects will
be negligible. For typical values ($Y=100\,{\rm MPa}$, $g=10\,{\rm m/s}^{2}$, $\rho_0=1.5\,{\rm g/cm}^{3}$,
$R=0.3\,{\rm mm}$, $\kappa_{z}^{eig}=1\,{\rm m}^{-1}$) the size of the unconfined
fiber ends $l_{conf}\approx2.5\,{\rm cm}$ is shorter than the 
fiber lengths of $l=10-20\,{\rm cm}$ typically used in our experiments, 
yet not entirely negligible. 
This is also the probable reason for filaments shorter than few cm being unable to
move and for sometimes occurring oscillatory behavior of the velocities of longer
filaments. For simplicity, in the following we will ignore these effects
assuming the fibers to be perfectly confined.}
confining the fiber to the interface plane -- in the two
macroscopic examples presented in Fig.~\ref{fig_Fiberboids-in-action},
the fiber is confined simply by its own weight. 
Given the confinement,
a fiber that is constantly driven by a flux normal to the confinement plane 
is mechanically frustrated\cite{PhysJournal}. Being unable to simply leave the plane, 
it follows an alternative path of least action and tends to axially rotate
by $\sim90^\circ$, so that it can realign its preferred curvature
with the x-direction. 

This results in the build-up of an in-plane curvature 
\begin{equation}
\kappa_{x}^{eig}\sim R^{-1}\Delta_{\leftrightarrow}\varepsilon_{yy}^{eig}, 
\end{equation}
where $\Delta_{\leftrightarrow}$ now stands for the left-right difference
(in-plane, perpendicular to the fiber axis).
Since the eigenstrain is proportional to the linear expansion coefficient $\alpha$, 
fiberboids can be divided into two classes:
``vexers'' having $\alpha>0$ and ``cavers'' having $\alpha<0$
(implying convex and concave shapes, respectively) as shown in
the snapshots in Fig.~\ref{fig_Sketch}b). 

The frustration-induced in-plane curvature 
is the precursor for the occurrence of the torque driving fiberboid rolling: 
the constant phase difference between the propellant-induced
strain (normal to the plane) and the realized strain (in-plane) 
results in a driving torque that induces a rotation of every cross-section 
with frequency $\omega$.
The driving torque density (per unit length) can be
calculated from slender rod theory, as demonstrated in appendix A.
By scaling, it is given as the product of the 
geometrically induced strain $\Delta_{\leftrightarrow}\varepsilon_{yy}^{eig}$
and the thermally induced stress $Y\Delta_{\updownarrow}\varepsilon_{yy}^{eig}$,
with $Y$ Young's modulus, integrated over the cross-section (i.e. times $R^{2}$),
yielding
\begin{equation}\label{driving_torque}
m_{drive}\sim R^2 Y 
\left(\Delta_{\leftrightarrow}\varepsilon_{yy}^{eig}\right)\left(\Delta_{\updownarrow}\varepsilon_{yy}^{eig}\right).
\end{equation}
Since the eigenstrain and the expansion coefficient $\alpha$ are linearly related
and $m_{drive}$ is quadratic in strain, the torque is insensitive to the sign of $\alpha$.
Together with the sign sensitivity of the in-plane curvature $\kappa_{x}^{eig}$, 
this explains the opposite relations of fiber curvature
and direction of motion for vexers and cavers. 
 
\subsection{Minimal model} \label{minimal model}

A minimal dynamic model, capturing the physical essence of most fiberboid
phenomena has three variables: the two non-dimensional eigenstrain
gradients $x=R\kappa_{x}^{eig},z=R\kappa_{z}^{eig}$ and the angular
frequency $\omega$:
\begin{eqnarray}
\dot{x} & = & -r\:x-\omega\:z\label{eq: x Dyn}\\
\dot{z} & = & -r\:z+\omega\:x+p\label{eq: z Dyn}\\
\omega & = & -\mu\:xz\label{eq: omega Dyn}
\end{eqnarray}
with constants $\left[r\right],\left[p\right],\left[\mu\right]=s^{-1}$
having dimension of a rate. The first two equations describe the
kinetics of the propellant-induced eigenstrains in the cross-section,
with $r$ and $p=\alpha q$ the strain relaxation and strain pumping
rates, respectively. The terms proportional to $\omega$ in Eqs.~(\ref{eq: x Dyn}-\ref{eq: z Dyn})
originate from the inter-conversion of the two modes via rotation.
Eq.~(\ref{eq: omega Dyn}) reflects the torque balance between a, 
here assumed Stokes-like, dissipational torque $\propto-\xi\omega$, 
with $\xi$ a friction constant, and the driving torque (density) from Eq.~(\ref{driving_torque}),
which reads $R^2Yxz$. 
The ratio
\begin{equation}
\mu\sim\frac{YR^{2}}{\xi} 
\end{equation}
occurring in Eq.~(\ref{eq: omega Dyn}) 
can be interpreted as a ``mechanical mobility'' that captures the elastodynamic response rate of the fiber.
We note that equations Eqs.~(\ref{eq: x Dyn}-\ref{eq: z Dyn}), as well as the
functional dependence of the driving torque can also be derived directly from
the underlying equations of thermo-elasticity of a slender rod and thermal advection-diffusion\cite{Baumann}.

\begin{figure}
\centering
\includegraphics[scale=0.8]{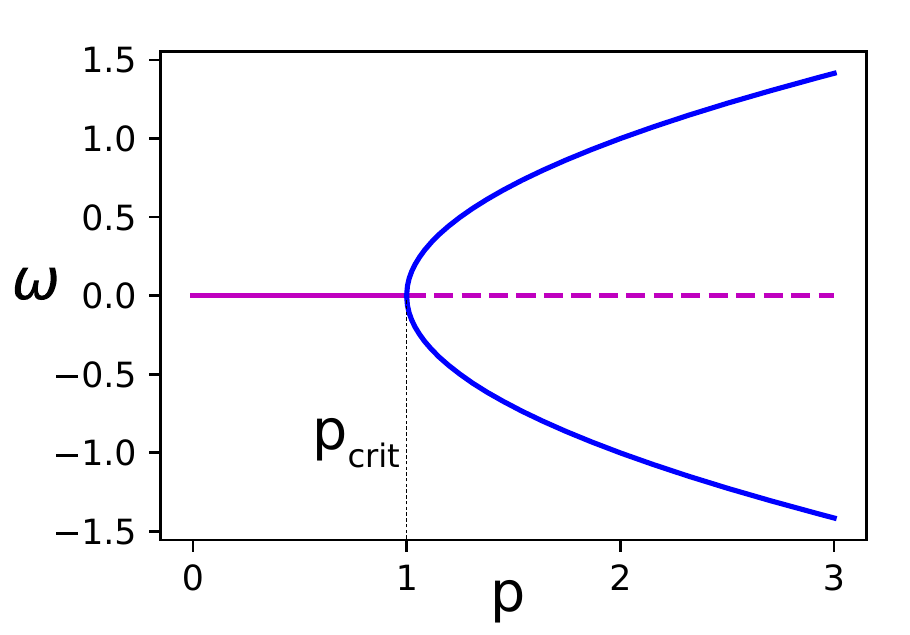}
\caption{\label{fig_bifurcation}
The steady state angular velocity $\omega$ of a fiberboid cross-section
as a function of the driving rate $p=\alpha q$, displaying a typical
supercritical pitchfork bifurcation. Parameters: $r,\mu=1$.
The trivial state $\omega=0$ looses stability at a critical driving strength
$p_{crit}$ and the direction of rolling (reflected by the sign of $\omega$)
is chosen by spontaneous symmetry breaking.}
\end{figure}

Let us study the steady states of the model.
First, obviously, an immobile state $\omega=0$ exists, with $x=0$ and $z=p/r$. 
In the case of thermal driving this state corresponds
to the thermal conduction state in absence of any advection. 
A second steady state emerges, provided the following relations hold
\begin{eqnarray}
\omega^{2}	&=&r^{2}\left(A^{1/2}-1\right)\label{eq:omega Steady}\\
x^{2}	&=&\frac{r}{\mu}\left(A^{1/2}-1\right)\label{eq:x Steady}
\end{eqnarray}
where $A$ is the (dimensionless) \textbf{activity parameter} given by
\begin{equation}
A=\frac{\mu p^{2}}{r^{3}}\label{eq:ActivityParameter}
\end{equation}
For the motile state to exist, the r.h.s.~of Eqs.~(\ref{eq:omega Steady}) and (\ref{eq:x Steady}) have to
be positive. This requires that the activity has to be larger than unity, 
$A>1$, implying a supercritical pumping rate
\begin{equation}
\left|p\right|>p_{crit}=\mu^{-1/2}r^{3/2}.
\end{equation}
Physically, the latter relation states that the pumping rate has to be larger than a threshold
value, namely large compared to the two forms of losses: 
$r$ (thermodynamic loss due to propellant relaxation) 
and $\mu^{-1}$ (frictional loss). 
The other way round, motion will stop, if the mobility $\mu$ 
drops below a critical value $\mu_{crit}=r^{3}p^{-2}$,
e.g.~in case when the friction increases due to a change of environment.
Note that while the translational velocity $v=R\omega$ always grows with $p$
and $\mu$, the in-plane curvature $\kappa=\frac{x}{R}$ is non-monotonous
in $\mu$ and vanishing both for small $\mu\le\mu_{crit}$ 
and large mobilities $\mu\to\infty$. 
A maximum curvature $\kappa^{*}=\frac{\left|p\right|}{2\,r}R^{-1}$
is attained at an intermediate mobility value $\mu^{*}=4\mu_{crit}$.

The model hence predicts a pitchfork bifurcation in the (angular) velocity 
at $p=p_{crit}$, cf.~Fig.~\ref{fig_bifurcation} . 
The rolling direction is spontaneously chosen, as observed for both the 
thermal\cite{Baumann} and the hygroscopic fiberboids.
Fiberboids are hence bidirectional motors and the onset of motion 
occurs via a spontaneous symmetry breaking.
It should be noted that in Ref.~\cite{Baumann},
dry friction instead of Stokes friction was used in the model, 
which renders the onset of motion discontinuous.

To experimentally probe the model predictions, Eqs.~(\ref{eq:omega Steady})
and (\ref{eq:x Steady}) can be used to obtain the following steady
state relation\footnote{
Divide Eq.~(\ref{eq:x Steady}) by Eq.~(\ref{eq:omega Steady}) to obtain $|x|=\frac{|\omega|}{\sqrt{\mu r}}$,
then use Eq.~(\ref{eq:omega Steady}) to get $\omega^2+r^2=r^2A^{1/2}=\sqrt{\mu r}p$, insert
and use $|x|=R|\kappa|$.}, 
\begin{equation}
|\kappa|=\frac{p|\omega|/R}{\omega^{2}+r^{2}},\label{eq: Master}
\end{equation}
connecting the two directly measurable variables, curvature $\kappa$ and angular velocity $\omega$.
It should be noted that the validity of this relation is independent
of $\mu$ and hence of the frictional mechanism --  
i.e.~the same relation can be obtained if dry friction is assumed in Eq.~\ref{eq: omega Dyn}.

\begin{figure*}[t]
\centering
\includegraphics[width=17cm]{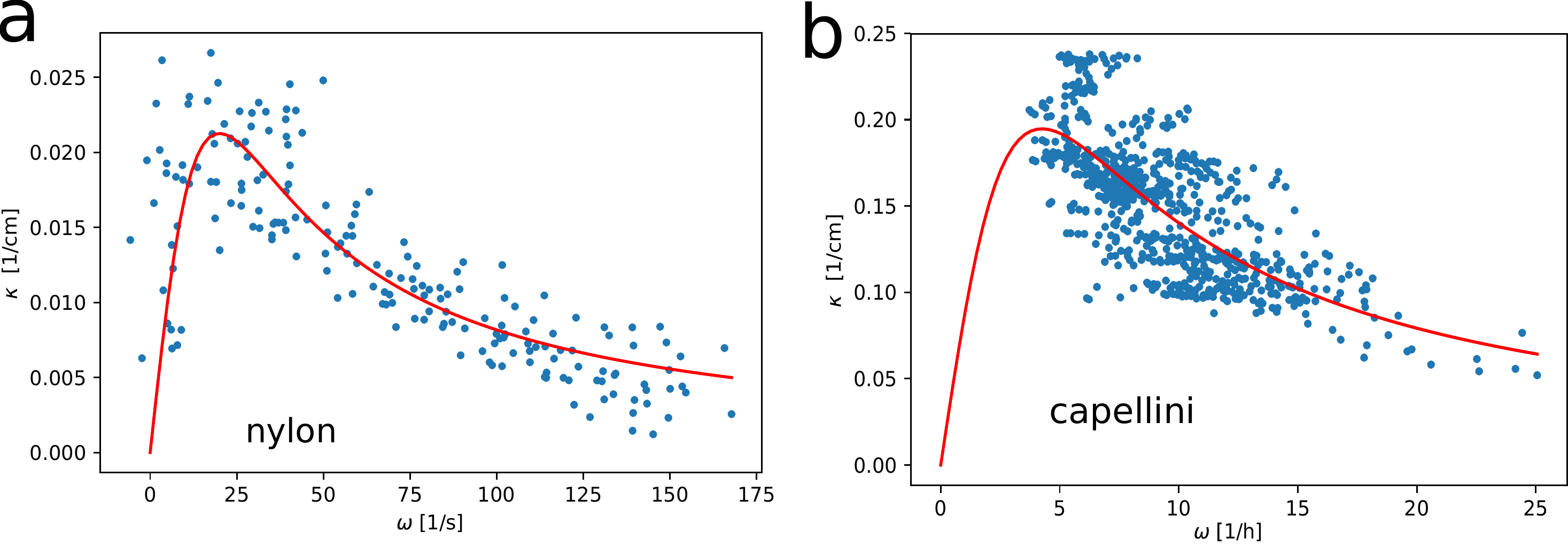}
\caption{\label{fig_mastercurv}
Master curves, as described by Eq.~(\ref{eq: Master}),
of curvature $\kappa$ vs.~angular velocity $\omega$ 
for nylon fibers on a heated plate (a), 
and thin spaghetti (capellini) on a wet towel (b). 
Blue dots are experimental data obtained from both transient motion and of steady
rolling fibers, the red curve is a fit to the theory, allowing to extract
values for the propellant pumping rates and the propellant relaxation rates:
one obtains $p=0.85\,[\pm 0.04$ (95\% confidence interval)]$\,{\rm s}^{-1}$ and $r=20.0\,[\pm 1.0]\,{\rm s}^{-1}$ for nylon (diameter $0.4\,{\rm mm}$, temperature $160^\circ$C)
and $p=1.66 \,[\pm 0.02]$  $\,{\rm h}^{-1}$ and $r=4.25\,[\pm 0.12]\,{\rm h}^{-1}$ for capellini (diameter $1\,{\rm mm}$, humidity 60\%, temperature $25^\circ$C.)}
\end{figure*}

Fig.~\ref{fig_mastercurv} shows experimental results for
both thermally driven nylon fibers in (a) 
and humidity-driven thin spaghetti (capellini) in (b). 
On the one hand, all the measured points, obtained both from
transient dynamics and steady rolling, fall on the master curve given
by Eq.~\ref{eq: Master}. This hence allows to extract the pumping and relaxation rates 
$p$ and $r$, as given in the figure caption.

\begin{figure}[b!]
\centering
\includegraphics[scale=0.55]{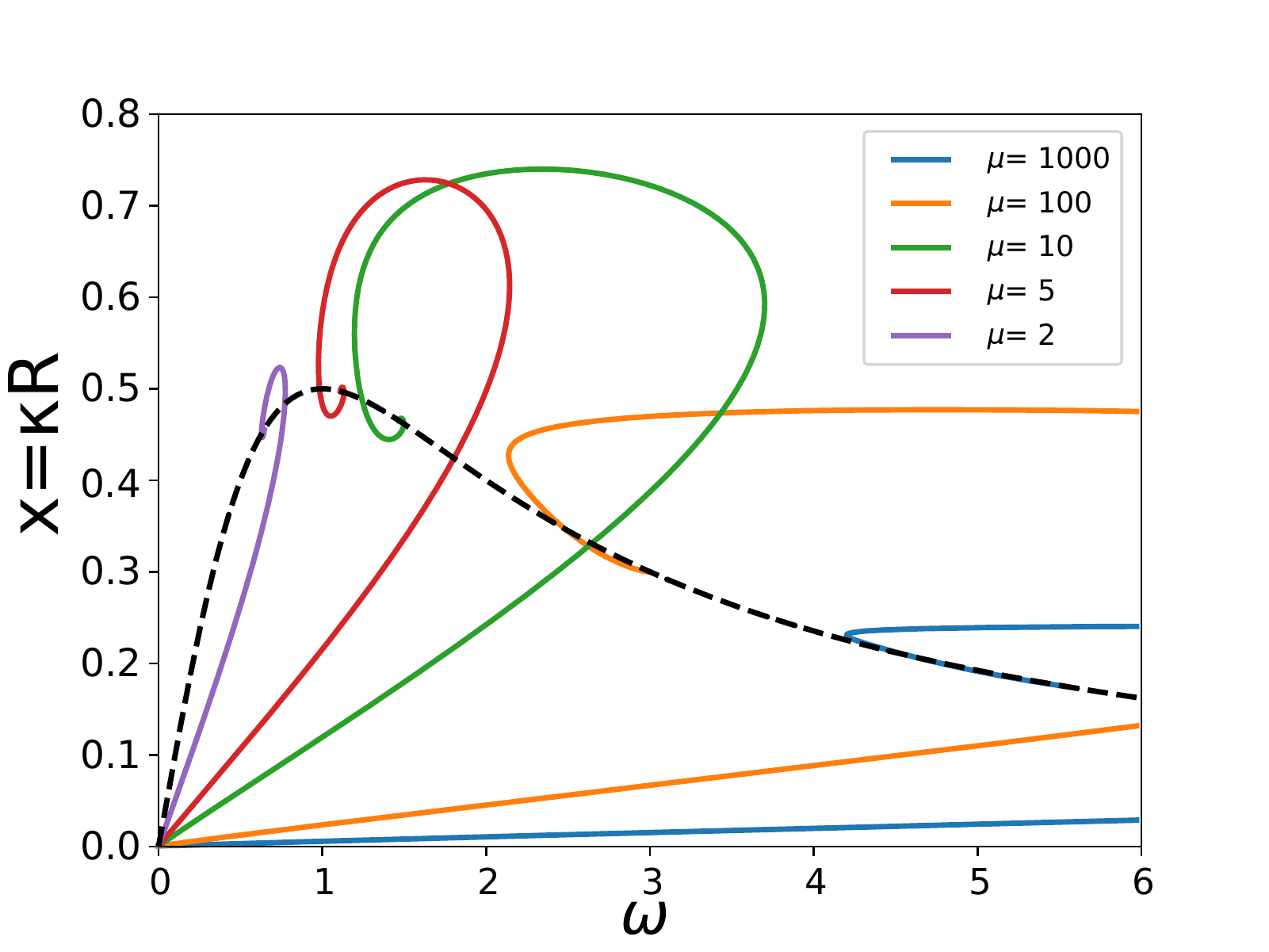}
\caption{\label{fig_trajectories}
Shown are numerically obtained  trajectories from Eqs.~(\ref{eq: x Dyn}-\ref{eq: omega Dyn})
in the plane in-plane eigenstrain $x\left(t\right)$ vs.~angular velocity $\omega\left(t\right)$
and the corresponding master curve for the stationary state.
All trajectories were initiated with zero strain (and hence curvature) and velocity,
$x\left(0\right)=0$, $\omega\left(0\right)=0$.
After a transient excursion they all land on the master curve, with
low mobility values $\mu$ corresponding to the ascending branch 
and higher mobilities to the descending branch of the master curve.
The black dashed line is the master curve given by Eq.~(\ref{eq: Master}) 
with $x=\kappa R$. Parameters: $r=p=1$.
}
\end{figure}

On the other hand, an interesting question arises: why do the points, 
even when obtained for a single sample, not condense into a single spot? 
In fact, a given mechanical mobility $\mu$ should select a single point 
on the master curve. This can be evidenced as in Fig.~\ref{fig_trajectories},
where numerically obtained  trajectories from Eqs.~(\ref{eq: x Dyn}-\ref{eq: omega Dyn})
are shown in the plane in-plane eigenstrain vs.~angular velocity.
All trajectories were initiated with zero strain (and hence curvature) and velocity.
After a transient excursion they all land on the master curve, with
low mobility values $\mu$ corresponding to the ascending branch 
and higher mobilities to the descending branch of the master curve.

\begin{figure*}[t!]
	\centering
	\includegraphics[width=17cm]{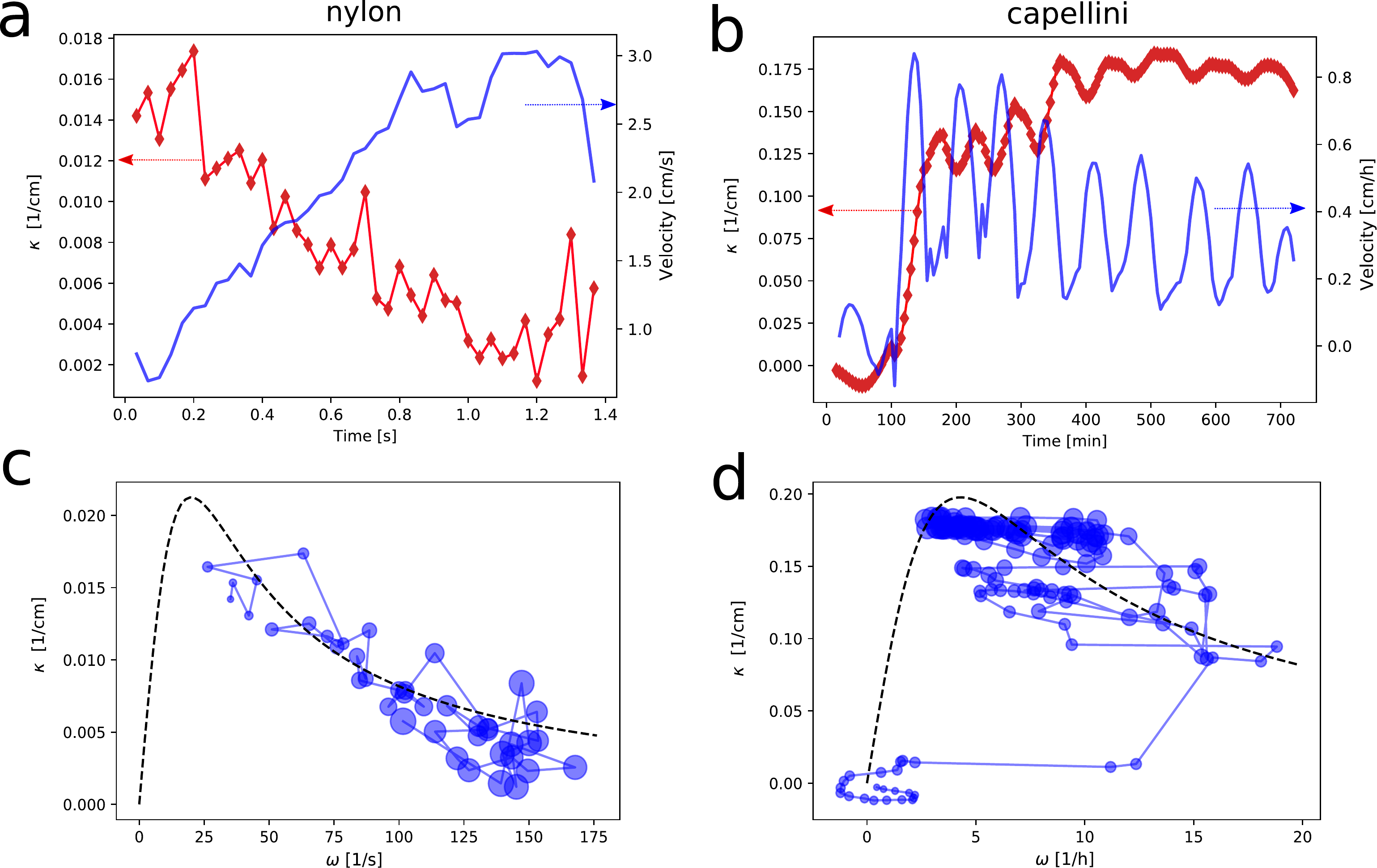}
	\caption{\label{fig_timedep_curv_vel}
		a), b) Time dependence of fiber curvature (red) and translational velocity (blue) 
		close to the onset of motion of fiberboids. 
		Part a) shows representative data for a thermally driven nylon fiber 
		and b) for a humidity-driven spaghetti, respectively.
		One can see that details are very different, especially there are transient oscillations
		for the noodles. Overall there is a certain anti-correlation between curvature and velocity,
		meaning the transients correspond to the descending branches of the respective 
		master curves. 
		This is evident when plotting the same data in the plane curvature vs.~angular
			frequency as shown in parts c) and d), where the dashed curves are the respective master curves.
	}
\end{figure*}

The experimentally observed variance along the master curve could be partially rationalized
by the inherent disorder of each sample (especially pronounced for
the starch-based noodles used) and the lack of end-piece confinement of
the filaments. The latter sometimes leads to a visible ``hinging''-type of
movement with a pronounced anti-phase oscillation of velocity and
curvature (see also Fig.~\ref{fig_timedep_curv_vel}b). 
This behavior can be interpreted in terms of a time-dependent
mobility $\mu\left(t\right)$ that oscillates during one rotation cycle. 
Furthermore, inhomogeneities of the (here rigid) substrate surface
can further contribute to variations in $\mu$. 
Both effects combined can lead to a ``migration'' along the $\kappa-\omega$ master curve 
for each sample.

\begin{figure}[b!]
\centering
\includegraphics[scale=0.55]{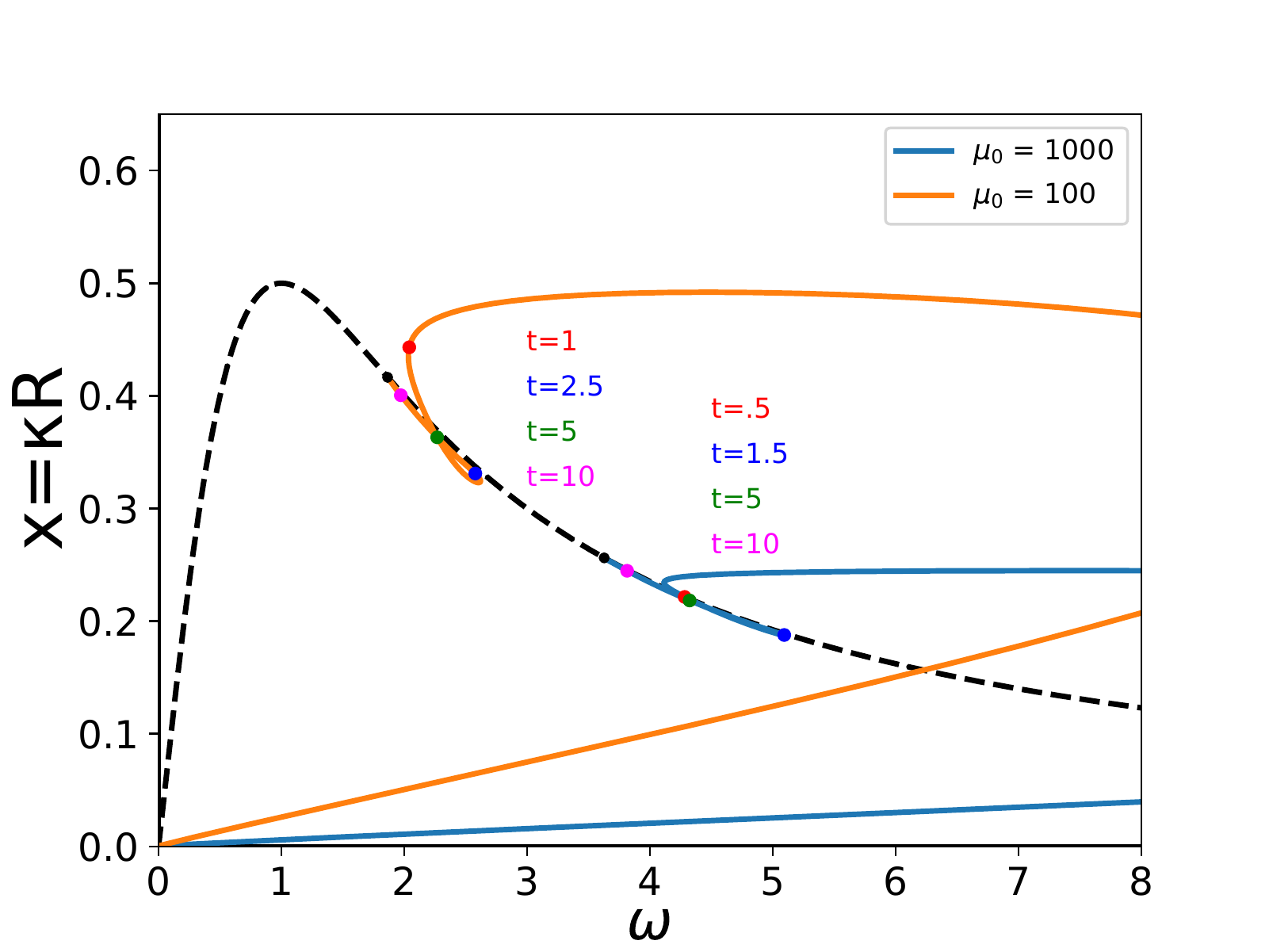}
\caption{\label{fig_trajYofpsi}
Shown are numerically obtained  trajectories from
Eqs.~(\ref{eq: x Dyn}-\ref{eq: omega Dyn}) as in Fig.~\ref{fig_trajectories}, but
accounting for the dynamics of the storage mode, Eq.~(\ref{eq: storage}).
We assumed $\mu=\mu_0(1-0.2\bar\psi)$ and $r_\psi=p_\psi=0.3$.
Hence the time scale for the dynamics of the storage mode is slower than the $x,z$-modes 
by roughly a factor of three, and the Young's modulus (and hence the mobility) is reduced
due the increased temperature to $20\%$ of its initial value (at room temperature). 
The black dashed line is the master curve given by Eq.~(\ref{eq: Master}) 
with $x=\kappa R$. One can see that for both examples, after a rapid transient with high
frequency and curvature excursions, the system stays close to the master curve
for a substantial amount of time and moves in a non-monotonic fashion
(see indicated times as colored dots).
}
\end{figure}

However, in addition to such intrinsic disorder effects, 
there are more systematic reasons for this behavior 
necessitating extensions of the minimal model, discussed in the next section.

\subsection{Refined relaxation models}
For the sake of simplicity, in the minimal model we have neglected
certain processes and materials behavior. First, besides $x$ and
$z$, there is a third dynamic ``mode'' that can become important: 
namely, a storage or internal reservoir mode, related to the average amount 
of propellant density, $\bar{\psi}$. 
The presence of this mode can be visualized experimentally by placing
a hot fiber (formerly rolling on a heated plate) on a room temperature
cold plate\cite{Baumann}. The fiber will still roll transiently -- until it has equilibrated
with the plate -- but now curves oppositely with respect to the direction of
motion as before, see supplementary movie 3. The reason is that, now, heat is flowing out of the fiber.

A second effect concerns the material properties of the fiber. 
In case of thermal driving, the storage mode corresponds to the average temperature in
the fiber cross-section and as such will affect the fiber's mechanical properties:
most importantly, the bending rigidity, determined by Young's modulus
$Y$, will vary with the average temperature (or the average density of propellant
in the general case). 
In addition, other material parameters may also vary, 
for instance the (linear) thermal expansion coefficient
depends on the reference temperature.

The storage mode, that was neglected in Ref.~\cite{Baumann}, can be included in the model on a generic level,
as well as directly derived from the underlying thermal diffusion-advection equation, see Appendix B.
Its dynamics is given by
\begin{equation}
\dot{\bar{\psi}}=\textrm{ }-r_{\psi}\bar{\psi}+p_{\psi},\label{eq: storage}
\end{equation}
hence analogously to the $z$-mode, the storage mode is pumped and relaxes, 
albeit with different (but related) rates $r_{\psi},p_{\psi}$. 
Eqs.~(\ref{eq: x Dyn}) and (\ref{eq: omega Dyn}) remain unchanged, but now 
the mobility $\mu=\mu(\bar{\psi})$ is a function of $\bar{\psi}$, 
stemming from the $\psi-$dependence of Young's modulus $Y(\bar{\psi})$.
The main effect of the storage mode is exemplified in Fig.~\ref{fig_trajYofpsi},
where we assumed that $\bar\psi$ relaxes slower than $x,z$ and leads to a substantial
softening of the Young's modulus. One can see that 
the system stays close to the master curve for a substantial amount of time 
while moving along it in a non-monotonic fashion (see colored dots).

Yet another complexity stems from the fact that the propellant transfer
is typically non-symmetric. This effect is especially pronounced 
when there is one solid-solid and one solid-air interface on the two sides of the fiber,
giving rise to vastly different boundary conditions on the flux:
considering thermal flux, typically, the conductive heat flux at the solid-solid interface 
is much faster than the diffusion-advection dominated flux at the solid-air interface.
In that case it can be shown (see Appendix C) that Eq.~(\ref{eq: z Dyn}) gets an additional term, 
coupling directly to the storage mode dynamics,
\begin{equation}
\dot{z}=-r\:z+\omega\:x+p+\chi(-r_{\psi}\bar{\psi}+p_{\psi}),\label{eq: z_with_storage}
\end{equation}
with $\chi$ a coupling coefficient, see appendix for details.

In case of a symmetric propellant transfer, i.e.~perfect thermal contact at
bottom and top with the boundary condition $T_{ext|bottom}=T_{solid}$ and $T_{ext|top}=T_{air}$,
it can be shown 
by solving the rotationally advected heat equation 
in the fiber cross-section (see appendix B) that $\chi=0$.
In this simplest case, the mean propellant
density, $\bar{\psi}$, influences the dynamics only via slow changes
in the mechanical parameters, as described above.
In contrast, if a more realistic boundary condition is employed, 
namely $l_{th}\frac{d}{d\rho}T+(T-T_{ext})=0$ accounting for convective heat transport 
in the surrounding liquid (in our case air), $\chi$ has the following properties, see appendix C: 
$\chi$ is (i) finite, (ii) proportional to the thermal expansion coefficient
$\alpha$ and (iii) increasing with the thermal length scale $l_{th}$,
until saturating for large values of $l_{th}/R$.

Hence there are two important slow modes, not present in the minimal model:
first, the dynamics of $\bar{\psi}$, 
which is slower than the one of the modes $x$ and $z$ (for geometrical reasons, $r_{\psi}<r$ holds), 
influences the dynamics via slow changes
in the mechanical parameters. And second,
in the general, asymmetric flux case, $\bar{\psi}$
also induces a slow variation of the effective driving of the $z$-mode 
like $p_{eff}(t)=p+\chi(-r_{\psi}\bar{\psi}(t)+p_{\psi})$, cf.~Eq.~(\ref{eq: z_with_storage}). 
Inspecting Eq.~(\ref{eq: storage}), 
as $\bar{\psi}$ is equilibrating, $p_{eff}(t)$ exponentially relaxes towards $p$.

When placing a nylon fiberboid having room temperature on a heated
plate, one hence expects the following slow relaxations in time of
the parameters: the thermal expansion of nylon is typically non-linearly
increasing\cite{Baumann} in the relevant region $T\in[100-180]^\circ$C,
meaning it slowly increases with the mean temperature $\bar{\psi}$
until it reaches its stationary value. Hence the effective pumping
rate of the $z$-mode, $p_{eff}(t)$, increases due to its proportionality
in the expansion coefficient $\alpha$. However, as the driving flux is asymmetric, 
at the same time the pumping rate relaxes due to equilibration of $p_{eff}$
towards $p$. Altogether, the behavior of the pumping rate is hence non-monotonous in time in the general case.
In contrast, the second material coefficient affected by the storage
mode, the mobility $\mu$, always decreases in time: it is proportional
to Young's modulus $Y$, which softens with temperature.

Fig.\ref{fig_timedep_curv_vel}c),d) show the time dependence of curvature
(red) and rolling velocity (blue) for a thermally driven nylon and 
a humidity-driven spaghetti fiberboid, respectively. In both cases,
one can observe an anti-correlation between curvature and velocity,
i.e.~the shown transient events close to the onset of motion 
are located on the descending branches of the respective master curves shown 
in Fig.\ref{fig_mastercurv}a),b).
Interestingly, for nylon curvature decreases while velocity increases
while noodles show the opposite trend, including oscillations related
to the less strict confinement of the fiber ends.

	\begin{figure}[h!]
	\centering
	\includegraphics[scale=0.9]{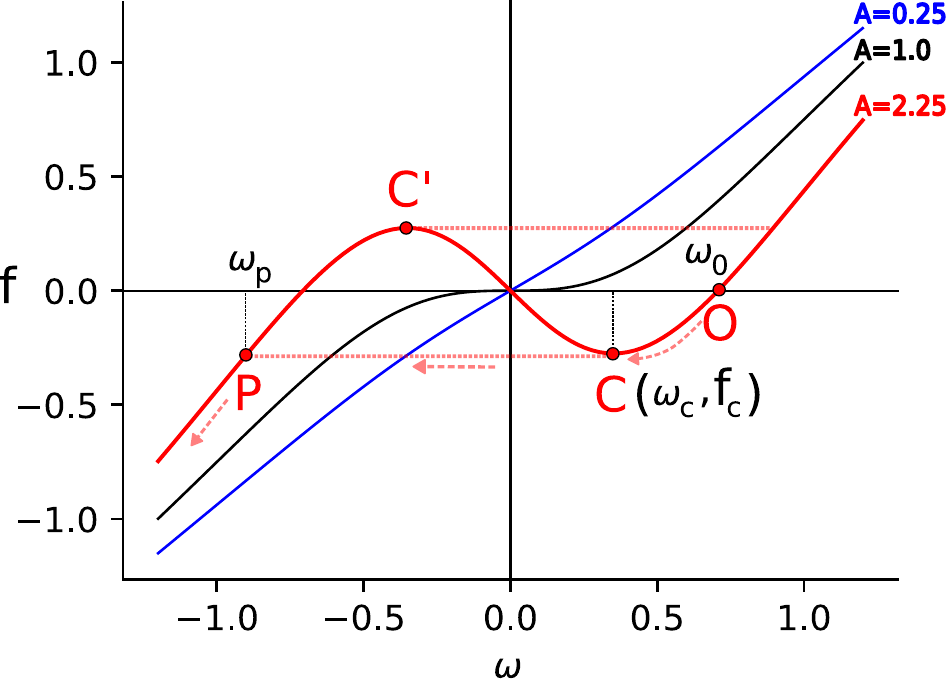}
	\caption{\label{fig_foce_velocity}
	The "motor relation" curves of fiberboids: scaled torque $f$ vs.~angular velocity $\omega$ 
	(in units of $r$), as given by Eq.~(\ref{eq:force-vel}) for selected values 
	of the activity parameter $A$ ($\mu=1$ , $r=1$, $p=0.5,1,1.5$). 
	For large activity $A>1$ beyond unity (red curve) the motor curve has a 
	typical hysteresis loop. Starting from zero torque $f=0$ (point O), 
	increasing the magnitude of the opposing torque ($f<0$) we reach the critical point C 
	with a critical torque $f_c$. Beyond $f_c$ the motion abruptly inverts direction 
	and $\omega$ drops to a large negative value $\omega_P<0$ following now the direction 
	of the driving torque (point P and further left to it). The branch $C C'$ is unstable.   
	}
	\end{figure}

\subsection{External forces and  torque-velocity relation} \label{external pot}

Up to now we have modeled fiberboids in uniform external environments and in absence of forces. 
To understand how they interact with external fields, 
substrate inhomogeneities or obstacles we can generalize the minimal model 
from Section \ref{minimal model} to motion in an external potential $V(\phi)$ 
acting on the cross-section rotation angle $\phi(t)=\int \omega ~ dt=S/ R$ 
(with $S$ the distance traveled on the substrate). 
Substituting $\omega=\dot{\phi}$ in Eq.~(\ref{eq: x Dyn}) and Eq.~(\ref{eq: z Dyn}) 
and adding an additional torque $V'(\phi)$ to Eq.~(\ref{eq: omega Dyn}) 
we obtain
\begin{eqnarray}
\dot{x} & = & -r\:x-\dot{\phi}\:z\label{eq: x Dyn_ext}\\
\dot{z} & = & -r\:z+\dot{\phi}\:x+p\label{eq: z Dyn_ext}\\
\dot{\phi} & = & -\mu\:xz+f(\phi)\label{eq: omega Dyn_ext}
\end{eqnarray}
with 
$f(\phi)=-V'(\phi)/\xi$.

\textbf{Torque-angular velocity relation.} 
Let us consider the simplest case where the external potential 
has a uniform gradient giving rise to a constant torque (scaled by friction) $f= const$. 
In this case we can consider the fiberboid as an engine working against the constant 
external load $f$. In the steady-state limit (i.e.~$\dot{\phi}=\omega$ a constant, 
$\dot{x}=\dot{z}=0$) it is easy to derive the (torque-angular velocity) "motor-relation" 
for fiberboids:
\begin{equation}
f\left(\omega\right)=\left[1-A\left(1+\frac{\omega^{2}}{r^{2}}\right)^{-2}\right]\cdot\omega \label{eq:force-vel}
\end{equation}
with the reoccurring activity parameter $A=\mu p^{2}/r^{3}$ 
introduced earlier in Eq.~\ref{eq:ActivityParameter} for the torque-free case. 
Depending on the magnitude of $A$, the torque-velocity curve can have one of 
two characteristic shapes, see Fig.~\ref{fig_foce_velocity}. In the "weakly active" case, $A<1$, 
for which autonomous motility is absent at vanishing torque,  
the motor-relation is monotonic, with the forcing and motion always pointing 
in the same direction. In this case the motor  exhibits a merely dissipative 
response with an effective internal friction $\xi_{eff}\propto f(\omega)/\omega$ that 
is positive, similar to the behavior of a fully passive object under the action of a torque. 

In the more interesting "strongly active" case, $A>1$, a non-monotonicity arises 
and for $-\omega_c<\omega<\omega_c$ the motor curve has negative slope. 
In this case the system exhibits effectively a negative internal friction 
and the motor can roll against the torque. 
This opposition to the external torque is only possible as long as the motor 
is not overloaded, i.e.~for $|f|<f_c$ below a critical torque $f_c$. 
Surpassing this torque, the motor inverts direction and runs with the torque.  
Close to the onset of motion, for $A\gtrsim1$  just above its threshold value, 
$\omega$ is small. This allows us to expand Eq.~\ref{eq:force-vel} and 
imposing $\partial f /\partial \omega=0 $ one obtains an estimate for the critical torque for $A\approx 1$:  
\begin{equation}
f_{c}\approx-\frac{2}{3\sqrt{6}}\,r\,\left(A-1\right)^{3/2}\text{  (for \ensuremath{A\rightarrow 1^{+}})}. \label{eq:f_c_A_1}
\end{equation} 
Along similar lines in the opposite limit of large $A$ we have

\begin{equation}
f_{c}\approx-\frac{3\sqrt{3}}{16}~r~A\text{ (for \ensuremath{A\rightarrow\infty})}\label{eq:f_c_large_A}
\end{equation}

The results are intuitive: a more strongly driven system, with larger $A$, 
has a larger magnitude of critical torque. At the threshold ($A=1$) the critical torque 
vanishes as the fiberboid is insufficiently driven and cannot bear any opposing load. 

	\begin{figure}[h!]
	\centering
	\includegraphics[scale=0.55]{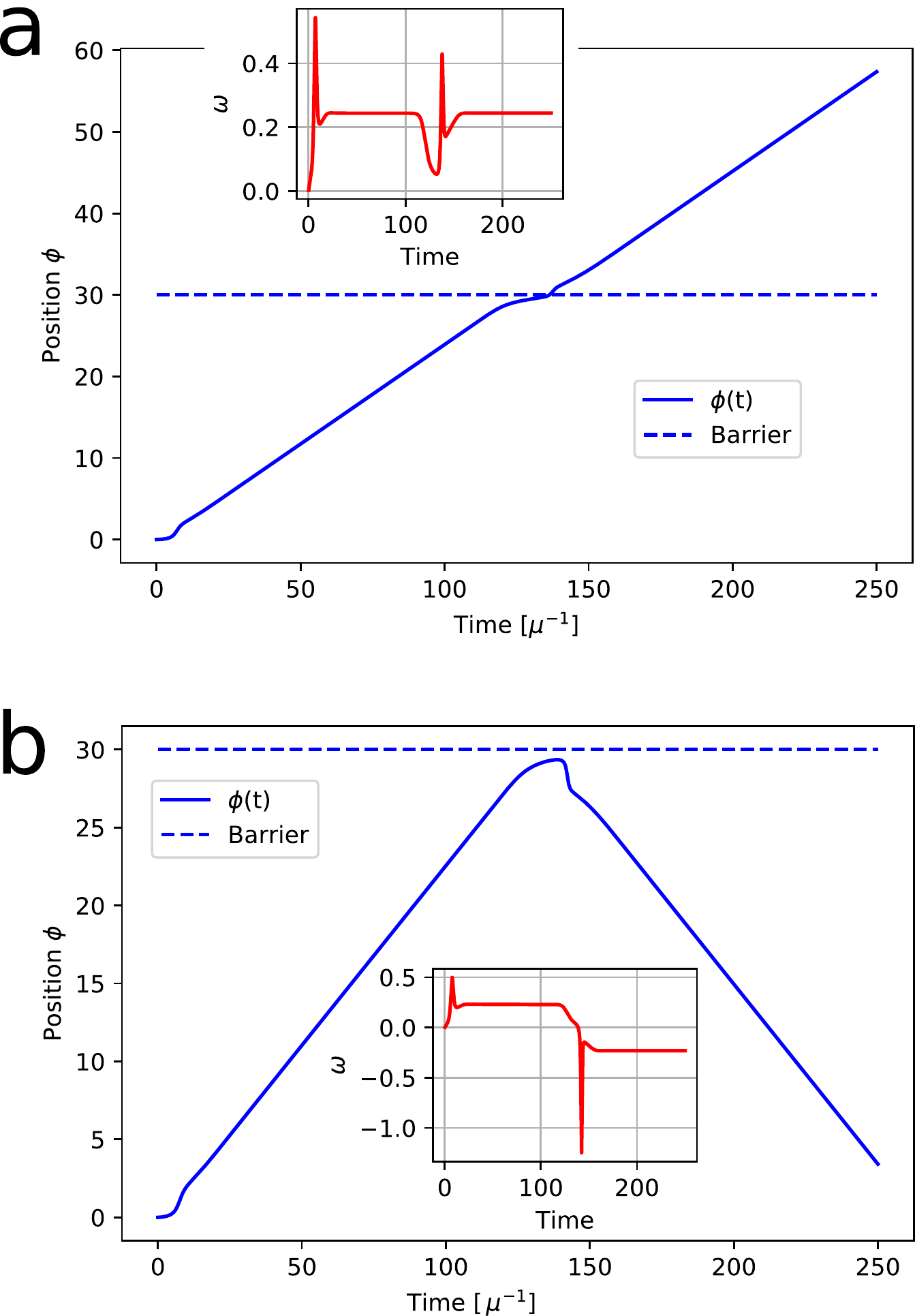}
	\caption{\label{fig_potential}
		    The dynamics of fiberboids interacting with an obstacle 
			(dashed line) as given by the dynamical Eqs.~(\ref{eq: x Dyn_ext}-\ref{eq: omega Dyn_ext}) 
			and the potential, Eq.~(\ref{eq: Potential}), with $V_0/s=1$,  $\mu=1$ and $r=0.1$. 
			Shown is the rolling angle (traveled distance/fiber radius)
			as a function of time. For supercritical driving ($p=0.22$, $A=48.4$) the fiberboid 
			overcomes the obstacle, see a),
			while for subcritical driving ($p=0.20$, $A=40$) it bounces off the obstacle 
			with a pronounced spike in angular velocity as shown in panel b).  
			Insets in a) and b) show the angular velocities (in units of $\mu$).
	}
\end{figure}

\textbf{Step barrier.} A second interesting case is that of a localized obstacle.
One can model an external obstacle at a position $\phi_0$ 
by a simple, smoothened step potential of the form
\begin{equation}
 V(\phi) = V_0 ~ \tanh\frac{\phi-\phi_0}{s} \label{eq: Potential}
\end{equation}
with the barrier height $V_0$ and $s$ 
a characteristic interaction range. 

Fiberboids hitting such an obstacle can behave in two characteristic ways: 
For strong drive (i.e.~$A$ large) and a weak obstacle (i.e.~small $V_0/s$), 
its angular velocity is only locally perturbed by the obstacle
and the fiberboid penetrates this ``barrier'' continuing in the same direction, 
see Fig.~\ref{fig_potential}a). 
In the regime of weak driving (small activity $A$) and sharper potentials (large $V_0/s$), 
however, the obstacle becomes effectively impenetrable and reflects the fiberboid, 
as shown in Fig.~\ref{fig_potential}b). 

Interestingly, in such a dynamic scenario of engaging a localized obstacle 
the fiberboids can be forced to invert direction by torques $|f|<|f_c|$ that are even below 
their steady-state critical value. In Fig.~\ref{fig_potential}a) and b), where the 
activities are $A=48.4$ and $A=40$, respectively, the predicted critical torques 
(from Eq.~\ref{eq:f_c_large_A}) are $f_c=-1.3$  and $f_c=-1.6$. 
Although the maximal torque exerted by the obstacle is in both cases $f=-1$, 
only the stronger driven fiberboid (the one in Fig.~\ref{fig_potential}a) 
can dynamically penetrate the barrier.  

\begin{figure*}
\centering
\includegraphics[width=17.5cm]{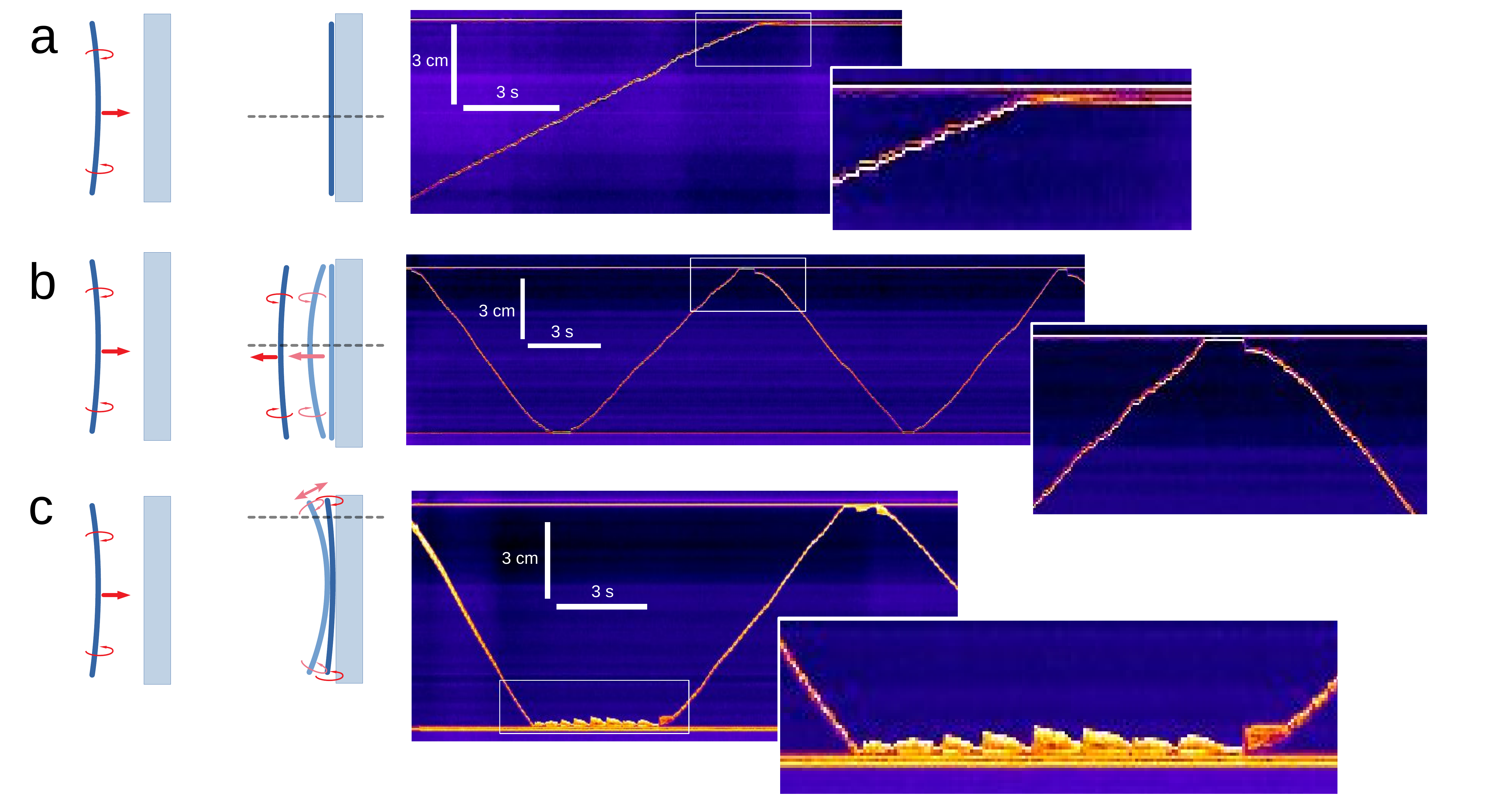}
\caption{\label{fig_collwall}
Typical outcomes for thermally driven nylon fiberboids 
hitting obstacles, implemented via glass slides lying on the heated plate. 
Panel a) shows the case for low driving rate ($T=130\protect^\circ$C), 
where the fiber gets stalled at the wall and completely stops rolling. 
Panel b) shows that for intermediate driving ($T=160\protect^\circ$C) 
the fiber is effectively reflected from the walls, in a nicely reproducible manner
(the kymograph shows 4 reflection events). Note the spike in velocity
immediately after bounce-off.
Panel c) shows that
for too strong driving ($T=170\protect^\circ$C),
the fiberboid gets captured at the wall, but continues rolling against it, 
performing a kind of flapping, cf.~the zoomed-in kymograph. 
The horizontal dashed lines in the sketches on the left indicate the position of the cross-section 
tracked in the respective kymographs on the right. 
See supplementary movie 5.}
\end{figure*}

Another interesting observation is that a reflection can happen with transiently 
highly elevated speeds, well in excess the steady state angular velocity. 
Such an accelerated bounce-off relaxation behavior is also observed 
experimentally, see the following Section and Fig.~\ref{fig_collwall}b). 
It can be rationalized by the observation that the magnitude of the steady-state angular velocity  
is always larger after the directional switching event than before it, $|\omega_p|>|\omega_c|$ 
(see the jump from point C to P in Fig.~\ref{fig_bifurcation}). The torque assists (hinders) 
the motion after (before) the direction inversion.

\section{``Social'' behavior: Interaction with walls and other fiberboids}\label{social}
As we have seen, fiberboids are bidirectional and display spontaneous symmetry breaking
at the onset of motion.  In the previous section we also proposed a 
phenomenological model for fiberboids bouncing into obstacles.
Both suggests that fiberboids should be able to respond 
to their environment, e.g.~by inverting their direction of motion,
and that they are poised to engage in interesting collective behavior, see supplementary movie 4. 
 We now study experimentally the two-body behavior and answer the questions: 
What happens when fiberboids engage immovable obstacles
or when they they frontally collide with each other?

The answer to this question depends on whether the fiberboids are
``vexers'' or ``cavers'', cf.~Fig.~\ref{fig_Sketch}b), 
i.e.~if they curve towards or away from
their direction of motion. In general, by simple geometric reasoning, 
cavers  (like thermally driven nylon) establish single-point contact at their center apex position,
while vexers (like humidity-driven noodles) make contact at their distant end points. 
The contact in the former case is hence much more stable and leads to more predictable interactions. 
In the latter case, the two-point contact renders the interaction
less predictable because the ends are often insufficiently confined to the plane.

\subsection{Cavers}
Fig.~\ref{fig_collwall} shows the generic scenarios found
for thermally driven fiberboids of ``caver''-type hitting a wall.
Experimentally, this can be simply studied by confining the rolling
fiber between two glass slides (placed as 2D obstacles) on the heating plate. 
Panel a) shows a weakly driven fiber ($T=130^\circ$C) that is simply
stopped by the wall. The reason is a combination of increased friction
due to the presence of the wall and unfavorable in-plane heat transfer
through the glass wall to the fiber. 
In fact, the back of a rolling nylon fiber is warmer than the front
(cf.~Fig.~\ref{fig_Sketch}a); for thermally driven fibers the color-code there 
means red=hot, blue=cold). But the front receives additional thermal
energy from the glass wall, disfavoring the motion. 

However, if the temperature is further increased, as shown in Fig.~\ref{fig_collwall}b)
for $T=160^\circ$C, the fiber gets effectively reflected from the
wall: the middle part of the fiber is slowing down and ``loosing''
curvature just to an extent that the fiber ends are able to catch up and
the thermal transfer through the glass can invert the direction of
motion. As seen in the same figure, reflections are nicely reproducable
at both walls and they are often accompanied by jumps
in velocity, cf.~Fig.~\ref{fig_potential}b).
For even higher driving, cf.~Fig.~\ref{fig_collwall}c), the fiber stubbornly and continuously
rolls against the wall, while its ends, not in direct contact to the
wall, perform a kind of flapping motion. 

At the contact with the wall several effects 
play a role making a quantitative theory very difficult.
(1) First, there is the opposing force exerted by the wall. 
(2) Second, there is a loss of the rolling contact (including slip at the bent sides of the fiber) 
and in addition an increased friction at the wall.
(3) The propellant (heat) transfer at the wall is severely altered:
the glass slides have similar temperature as the heated surface,
which should give rise to an additional pumping term (similar to $p$) 
in the equation for the planar strain $x$. In addition, the cooling effect
of air convection is altered by the presence of the wall (cf.~the discussion of propellant flux asymmetry and the thermal
length scale $l_{th}$ in appendix C).
Finally, due to curving, it is no longer possible to treat the cross-sections
along the fiber as equal -- some are in contact to the wall while others, especially the ends, are not.

Nevertheless, the models developed in section \ref{generic_model} 
can account phenomenologically for most of the generic aspects of fiberboids colliding 
with walls (or with other fiberboids). 
In fact, the case shown in Fig.~\ref{fig_collwall}a) corresponds
to weak driving (small $p$ but $A=\frac{\mu p^2}{r^3}>1$ in the absence of the wall) 
and a rather rough obstacle whose effect can be interpreted by a lowering of the mobility $\mu$
upon contact as a combination of effects (1) and (2) discussed above. 
The fiberboid stops since the first term in Eq.~\ref{eq:omega Steady}
decreases with the reduced $\mu$ resulting from wall friction:
$A$ becomes less than one and rolling is no longer possible.

The case of Fig.~\ref{fig_collwall}b) corresponds to intermediate $p$, hence $A$ is larger
and the reduction of mobility upon contact does not reduce $A$ below one. Consequently, a finite rotation speed is still possible. However, Eq.~\ref{eq:x Steady} implies 
that the curvature is still reduced due to friction: 
the first, positive contribution scales like $\mu^{-1/2}$ while the
second, negative contribution as $\mu^{-1}$. 
This leads to the catch-up of the ends described above and then the fiber 
is ready to invert due to the additional heat influx from the wall, cf.~point (3) above,
into what will become the new back of the reversed fiber. 
If $p$ is even larger, the reduction in $\mu$ is not sufficient to reverse 
the fiber and rolling against the wall persists, corresponding to the case 
shown in Fig.~\ref{fig_collwall}c).

The cases shown in  Fig.~\ref{fig_collwall} can be also interpreted 
using the model including an external potential as described in section \ref{external pot}.
Here the potential models the effects (1) and (2) from above on a phenomenological level.
Case b) corresponds to the case where the driving is intermediate, as shown in 
Fig.~\ref{fig_potential}b). The force-velocity relation in addition
nicely explains the pronounced spike in velocity seen immediately after the inversion of the fiber,
see Fig.~\ref{fig_collwall}b).
Lower driving corresponds to stopping, while larger drive to continuing rolling 
without inversion, where however the fiberboid cannot penetrate the step potential at all.
It is likely that the complete stopping at the wall is furthermore strongly affected by additional effects, here not taken into account, like the critical threshold for the onset of motion under dry friction.

\begin{figure*}
\centering
\includegraphics[width=18cm]{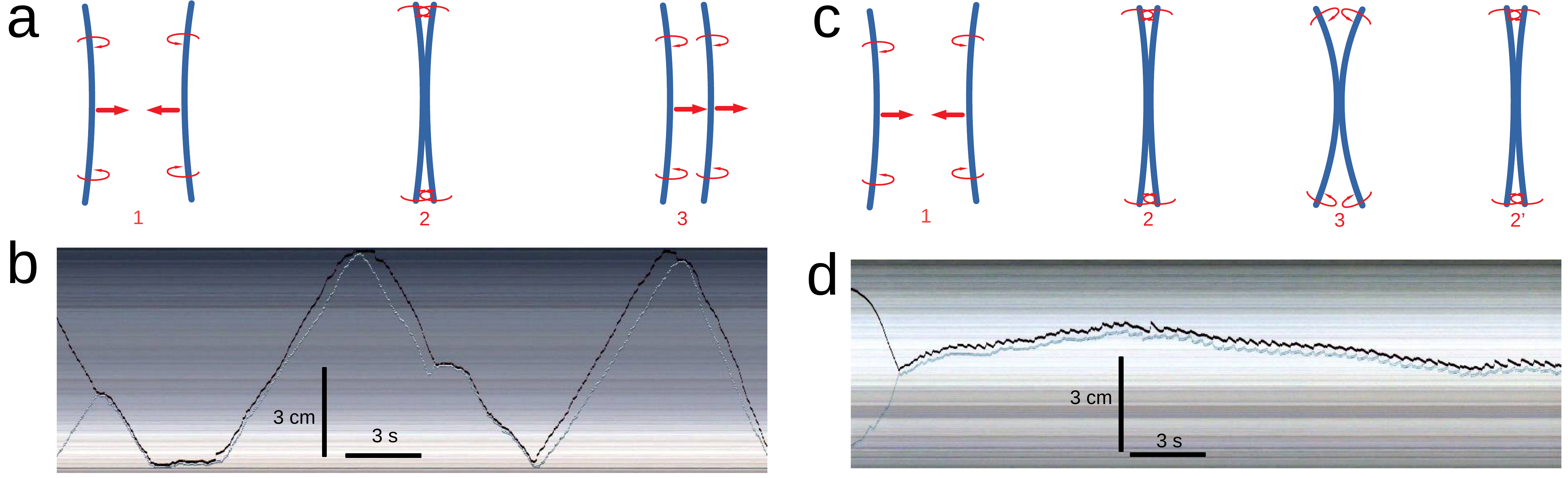}
\caption{\label{fig_bin_collisions}
Typical binary collision events of thermally driven nylon fiberboids of equal length.
Two main outcomes for the head-on collision dominate: 
panel a) sketches an event, where after rolling against each other for a
short transient time, one fiber switches directionality such that finally
both move together as a pair with almost equal velocity. 
This can be seen in the kymograph shown in b) at the very left. 
Note that the kymograph also displays several collisions of the fiberboid pair 
with boundaries, where the pair is effectively reflected. 
Hence this ``follower-leader'' mode is quite robust against perturbations. 
Panel c) sketches a second common event, where the pair keeps on rolling against each other, 
with the center of mass of the ``dancing pair'' slightly going back and
forth due to curvature variations (``flapping''). As can be seen
from the time scale bar in d), this state can be stable 
for a substantial time (minutes). 
} 
\end{figure*}

For binary collisions between two fiberboids, quite similar effects to wall collisions are observed. 
Fig.~\ref{fig_bin_collisions} shows the main types of events for binary central collisions
between two thermally driven fiberboids of equal lengths.
Panels a),b) show what we call ``follower-leader'' behavior: for a short transient,
both fibers roll against each other, but then the ``weaker'' inverts direction
and the pair now travels together. This pairing is quite robust against further
collisions or perturbations.
Panels c),d) shows what we call ``dancing pair'' behavior: here, 
the pair keeps on rolling against each other for long times.
At the same time, the common center of mass goes back and forth, 
due to flapping-type curvature variations.

\subsection{Vexers} 

\begin{figure}
\centering
\includegraphics[scale=0.4]{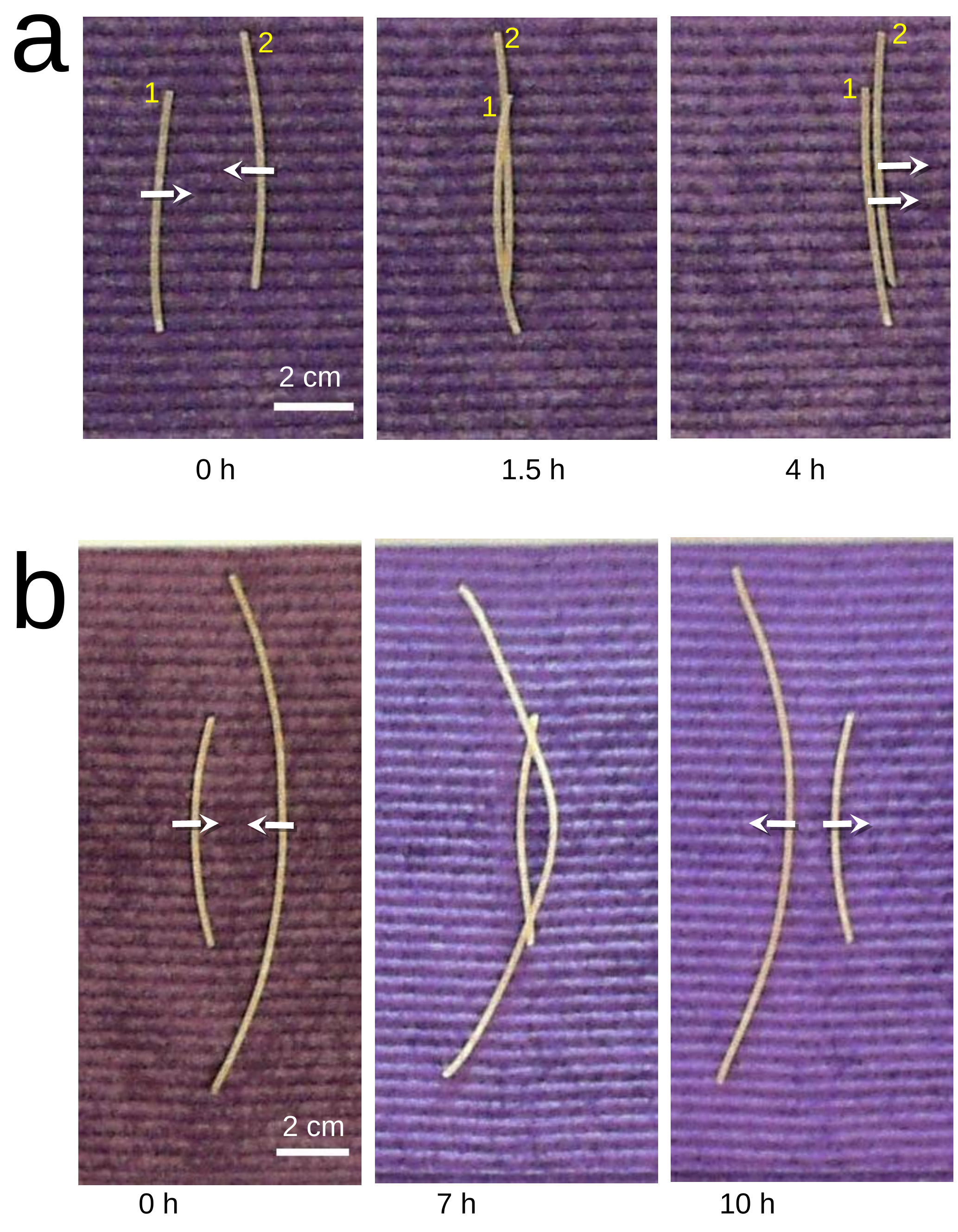}
\caption{\label{fig_collisions-noodle}
Typical binary collision events for vexers, realized by spaghetti on a wet towel. 
Panel a) shows a collision and inversion of short spaghetti,See supplementary movie 6, b) a collision 
of one long and one short spaghetti. In the latter case, the longer specimen 
transiently leaves the planar confinement and ``jumps'' over the shorter one. 
See supplementary movie 7. 
}
\end{figure}

Vexers, like the hygroscopically driven spaghetti, 
can exhibit similar ``social behavior'' as cavers,
as exemplified in Fig.~\ref{fig_collisions-noodle}a)
by an inversion event upon a collision.
However, unlike cavers, vexer-type fiberboids exhibit 
more random, generically unstable, multi-point contact configurations. 
This leads more frequently to complex out-of-plane deformations, 
followed by mutual entanglement or amusing ``tunneling'' behavior where one filament 
lifts and bypasses the other, as exemplified in Fig.~\ref{fig_collisions-noodle}b). 
Such richness and variability of collision outcomes makes vexers overall
less suitable for investigating their collective behavior.

\section{Conclusions and Outlook}

Fiberboids are a new member of the class of active, self-propelled particles\cite{ActiveParticles,SPProds}
that was surprisingly long hiding in the plain view, before being
only recently described\cite{Baumann}.

Fiberboids naturally share some similarities with other 
actively rolling systems, like Quincke spinners\cite{QuinckeLyon}, 
Marangoni-driven droplets \cite{Dauchotdroplets} or
Leidenfrost droplets\cite{Leidenfrost, Querewheel}. 
However, in contrast to those they do not rely on fluid motion.
They do develop a continuously propagating internally rotating
mode akin to B\'enard or Marangoni convection rolls in fluids, yet surprisingly this mode 
is an elastic prestress "flowing" through a solid material\cite{Baumann}.
This naturally has several advantages: the self-propelled particle is not eaten-up
by the propulsion process itself (as for Leidenfrost and Marangoni driving) and it can result,
especially for the thermal drive, in a constant steady-state velocity.

Another advantage of the fiberboid mechanism is its practical simplicity:
Many everyday objects can be easily turned into fiberboids, as demonstrated here: 
pieces of fishing line spin quickly on a hot plate, while spaghetti noodles 
spin both rapidly on a hot plate\cite{Baumann}, as well as much more slowly on wet cloth. 
The latter example also shows
that different fluxes can be employed to drive fiberboids, and on vastly different timescales
(differing by orders of magnitudes).
Last but not least, fiberboids are anisotropic objects and hence should display
interesting new collective effects beyond the existing spherical rollers \cite{QuinckeLyon,Dauchotdroplets,Leidenfrost, Querewheel}. 

Here we have laid out the basic phenomenology and proposed generic models
of increasing complexity for the dynamics of fiberboids. We started from
a simple two-mode model, describing the dynamics of the propellant gradients
in two orthogonal directions of the fiber cross section. 
This simple  model phenomenologically captures many of the 
features of fiberboid behavior, including the symmetry breaking at the bifurcation
point and a curvature-velocity dynamics following a simple master curve.
To treat all the experimentally observed phenomena, including dynamic transients, 
we extended the two-mode description and introduced a third degree of
freedom -- a ``storage mode'' corresponding to the mean propellant density 
in the cross section. Interestingly, this mode couples in several ways, 
namely both via the mechanics and the dynamics, to the other modes.

By studying fiberboid collision events, both with walls and other fiberboids
of the same type, we have extracted the 
conceptual ingredients for future studies of the collective behavior of (many) fiberboids. 
Although many of the generic aspects of the fiberboid collision events
could be rationalized by the model, others require future model sophistication and extensions. 
Purely 1D systems of interacting active particles have been studied 
for Marangoni droplets \cite{1D-Droplets} and crawling cells \cite{1D-Cells} 
in confined spaces. These also showed both velocity inversion and follower-leader behavior 
and, accordingly, complex collective behavior. 
Fiberboids add another, easy to study, table-top example 
to this generic class of phenomena able of unlocking the charms of non-equilibrium physics 
to undergraduate and high-school students, as we find from own experience.

We conclude with a simple message: fiberboids should be a general phenomenon
in nature, as they require only a few physical ingredients, which are:
a) An interface between two phases A and B, at least one of which
is a fluid (to allow for mobility). b) The phases A and B should be out-of-equilibrium, 
giving rise to a flux of energy/matter across their interface. 
c) Straight filaments, with cylindrical cross-section,  that are pinned (confined) at the interface
{\it and} that mechanically respond to the flux. 

These three conditions are indeed generic, also on the microscopic scale, 
as we are surrounded by interfaces that frequently are out-of-equilibrium. 
Naturally ordered, biological fibers can be trapped there by 
Van der Waals or electrostatic forces or Pickering pinning, and driven by a multitude of
chemical and thermodynamic fluxes. Moreover, it has been shown that many biofilaments, 
like microtubules\cite{MT1,MT2} and intermediate filaments\cite{IFCurvature} can spontaneously
break their cylindrical symmetry and self-buckle. This phenomenon happens under 
internal prestress and leads to symmetry broken zero energy (``wobbling'') modes. 
The latter zero modes are similar to the dissipatively emerging (rotating) curvature
mode in fiberboids\cite{Baumann} and should further facilitate the emergence of the
fiberboid effect by eliminating the threshold for buckling. Related
ideas of generating fluxes via various ratchet effects in circular
DNA\cite{DNA RIng} could further expand our toolbox to ``motorize any filament''.

Recently, we have speculated \cite{PhysJournal} that filamentous
viruses like ebola might utilize fiberboid physics to roll on surfaces.
Considering the ease with which we generated fiberboids, in particular
the motorization of spaghetti in various ways, 
it seems in fact unlikely that nature could have overlooked the motif.

\appendix

\section*{Appendix A: Driving torque}

Here we calculate the driving torque generated by the fiberboid mechanism
using slender rod elasticity coupled to the propellant density which induces
expansion/contraction according to Eq.~(\ref{eigenstrain}). 

Let the interface be the $X$-$Y$-plane. 
We define the lab frame by laying the fiber axis along $Y$
and consider a cross-section of the fiber with 
the fiber's (vectorial) curvature fixed along  $X$, hence $\bkap\cdot e_x=\kappa_x=\kappa$.
We introduce the angle $\Phi$ with respect to the $X$-axis, 
describing a rotation of the fiber's cross-section along its axis
(i.e.~the $Y$-axis) and defining the co-rotating internal coordinate system 
$e_1=(\cos\Phi,\sin\Phi)\,,\,\,\,e_2=(-\sin\Phi,\cos\Phi)\,$.
We call these internal coordinates $(X',Z')$ or in polar form $(\rho,\phi)$.
Obviously, they are related to the lab frame coordinates via a rotation matrix
\begin{equation}\label{rotmatrix}
\left(\begin{matrix}
       X' \\ Z'
      \end{matrix}\right)=
\left(\begin{matrix}
       \cos\Phi & \sin\Phi \\ 
       -\sin\Phi & \cos\Phi
      \end{matrix}\right)
\left(\begin{matrix}
       X \\ Z
      \end{matrix}\right).
\end{equation}

In these internal coordinates the axial strain in the cross-section ($\ep_{YY}$) 
for a slender rod is given by
\begin{equation}
\ep(\rho,\phi)=-\rho(\kappa_1\cos\phi+\kappa_2\sin\phi)+\bar\ep-\ep_\psi(\rho,\phi)\,.
\end{equation}
The first contribution is bending (with the negative sign indicating 
compression within the cross-section at the inwards curved side). 
$\bar\ep$ is the strain related to axial compression/elongation
and $\ep_\psi(\rho,\phi)$ is the propellant-induced (axial) prestrain. 
Using that the curvatures in the rotating frame are given by 
$\kappa_1=\bkap\cdot e_1=\kappa\cos\Phi\,,\,\,\,\kappa_2=\bkap\cdot e_2=-\kappa\sin\Phi$,
one gets 
\begin{equation}
\ep(\rho,\phi)=-\rho\kappa\cos(\phi+\Phi)+\bar\ep-\ep_\psi(\rho,\phi)\,.
\end{equation}

The elastic energy of the considered cross-section (i.e.~energy per length)
reads 
$E=\frac{Y}{2}\int\left( \ep(\rho,\phi) \right)^2\rho d\rho d\phi $
with $Y$ Young's modulus. Inserting the strain yields three contributions
\begin{eqnarray}\label{elen1}
E(\kappa,\bar\ep,\Phi)
&=&\frac{1}{2}B\kappa^2+\frac{Y}{2}\int\left(\bar\ep-\ep_\psi(\rho,\phi)\right)^2\rho d\rho d\phi\nonumber\\
&&+Y\kappa\int\cos(\phi+\Phi)\ep_\psi(\rho,\phi)\, \rho^2 d\rho d\phi.
\end{eqnarray}
Clearly, the first term the standard bending energy with
$B=\frac{\pi }{4}Y R^4$ the bending stiffness.
To evaluate the others, one needs an expression for $\ep_\psi$.

The propellant-induced (axial) prestrain, cf.~Eq.~(\ref{eigenstrain}),
is given by
$\ep_\psi(\rho,\phi)=\alpha\psi(\rho,\phi)$,
where complications such as shear in the cross-section, torsional contributions etc.~are disregarded
for simplicity.
For the propellant distribution we make the following ansatz,
\begin{equation}\label{Temp_ansatz_rot}
\psi(\rho,\phi)=\Delta \psi_c\frac{\rho}{R}\cos\phi+\Delta \psi_s\frac{\rho}{R}\sin\phi+\tilde{\psi}\,,
\end{equation}
introducing the two principal differences (`modes') in the rotating frame, 
$\Delta\psi_c$ and $\Delta\psi_s$, as well as the homogeneous contribution $\tilde{\psi}$.

We now assume that the compressive/elongational strain relaxes fast.
Minimization of the energy, Eq.~(\ref{elen1}), w.r.t.~$\bar\ep$ 
yields 
\begin{equation}\label{minep}
\bar\ep=\frac{1}{\pi R^2}\int\ep_\psi(\rho,\phi)\rho d\rho d\phi\,.
\end{equation}
Insertion of the ansatz, Eq.~(\ref{Temp_ansatz_rot}), then simply yields
$\bar\ep=\alpha\tilde{\psi}$,
meaning that the compressive/elongational strain just compensates 
the homogeneous part of the propellant-induced stress.

We use this to simplify the elastic energy, perform all integrals and get
\begin{equation}\label{Eeltherm}
E(\kappa,\Phi)=B\left\{\frac{1}{2}\kappa^2
+\alpha\frac{ \kappa}{R}(\Delta\psi_c\cos\Phi-\Delta\psi_s\sin\Phi)
\right\}\,,
\end{equation}
plus an additional term that is curvature- and angle-independent and hence 
has no consequence for the dynamics.

Neglecting initial, transient reshaping events of the linear fiber,
the stationary curvature $\kappa_s$ of the fiber can be obtained 
by minimization w.r.t.~$\kappa$, 
yielding
\begin{equation}
\kappa_{s}=- \frac{\alpha}{R}(\Delta\psi_c\cos\Phi-\Delta\psi_s\sin\Phi)\,.
\end{equation}
In turn, the driving torque, generating the rolling motion if it overcomes the prevalent dissipation
mechanisms, is the conjugate quantity to the angle $\Phi$ and hence 
given by $m=-\frac{\p E}{\p \Phi}$, implying
\begin{equation}
m=B\alpha\frac{\kappa}{R}  (\Delta T_c\sin\Phi+\Delta T_s\cos\Phi)\,.
\end{equation}
Both expressions involve the modes $\Delta\psi_c$ and $\Delta\psi_s$
in the rotating frame. To express them in lab frame
one uses the definition of polar coordinates and Eq.~(\ref{rotmatrix}) 
in the ansatz, Eq.~(\ref{Temp_ansatz_rot}), to get
\begin{eqnarray}\label{Tansatzlab}
\psi(\rho,\phi)-\tilde{\psi}\hspace{-2mm}&=&\hspace{-2mm}
\left(\Delta\psi_c\cos\Phi-\Delta\psi_s\sin\Phi\right)\frac{X}{R}
+\left(\Delta\psi_c\sin\Phi+\Delta\psi_s\cos\Phi\right)\frac{Z}{R}.\nonumber
\end{eqnarray}
The two principal modes in the lab frame are hence identified as
\begin{eqnarray}
 \Delta\psi_x&=&\Delta\psi_c\cos\Phi-\Delta\psi_s\sin\Phi=:\Delta_{\leftrightarrow}\psi,\nonumber\\
 \Delta\psi_z&=&\Delta\psi_c\sin\Phi+\Delta\psi_s\cos\Phi=:\Delta_{\updownarrow}\psi\,.
 \end{eqnarray}
This leads to the simple and instructive expressions
\begin{equation}
\kappa_{s}=- \frac{\alpha}{R}\Delta_{\leftrightarrow}\psi\,\,\,,\,\,\,\,\,
m=B\alpha\frac{\kappa_s}{R} \Delta_{\updownarrow}\psi.
\end{equation}
I.e.~the curvature is determined by the in-plane  propellant gradient and the 
driving torque by both this curvature and the out-of plane propellant gradient.
Putting both together and absorbing the expansion coefficients by introducing 
the eigenstrains yields
\begin{equation}
m_{drive}=- \frac{B}{R^2} 
\left(\Delta_{\leftrightarrow}\varepsilon_{yy}^{eig}\right)
\left(\Delta_{\updownarrow}\varepsilon_{yy}^{eig}\right),
\end{equation}
hence the same result obtained by scaling arguments in Eq.~(\ref{driving_torque})

\section*{Appendix B: Symmetric thermal drive}

In the case of thermal driving,
one has to solve the rotationally advected thermal diffusion equation 
\begin{equation}
\label{rescaled}
\p_t T=\frac{D}{R^2}\nabla^2 T +\omega\p_\phi T\,.
\end{equation}
Note that we rescaled space by the fiber radius $R$, such that in polar coordinates, $T(\rho,\phi)$,
$\rho\in[0,1]$. For the simplest case of symmetric thermal contact, the boundary condition (BC)
at the cylinder surface reads
\begin{equation}
\label{BC}
T(1,\phi)-T_{ext}(1,\phi)=0\,.
\end{equation}
To capture the essential features, this BC is approximated by
\begin{equation}\label{apprBC}
T_{ext}(1,\phi)=\frac{T_s-T_{air}}{2}\cos\phi+\frac{T_s+T_{air}}{2}
=T^-\cos\phi+T^+\,,
\end{equation}
Like that $T_{ext}(1,0)=T_s$ is the temperature at the bottom ('surface')
and $T_{ext}(1,\pi)=T_{air}$ is the temperature at the top ('air').

We now are looking for a solution of the form
\begin{equation}\label{ansatz0}
T=C(\rho)\cos\phi+S(\rho)\sin\phi+\tilde{T}(\rho)\,. 
\end{equation}
We will later identify the $z$-mode of the main text (as $C$), the $x$-mode (as $S$) and 
the storage mode $\bar\psi$ (as $\tilde{T}$), respectively.
The symmetry-adapted functions to the problem are the so-called Zernike polynomials
which have the general form $Z_n^m=R_n^m(\rho)\cos(m\phi)$, $Z_n^{-m}=R_n^m(\rho)\sin(m\phi)$.
We will need only the first few which are given by
\begin{eqnarray}
&&Z_1^1=2\rho\cos\phi\,,\,\,\,Z_1^{-1}=2\rho\sin\phi\,,\nonumber\\
&&Z_3^1=\sqrt{8}(3\rho^3-2\rho)\cos\phi\,,\,\,\,Z_3^{-1}=\sqrt{8}(3\rho^3-2\rho)\sin\phi\,,\nonumber\\
&&Z_0^0=1\,,\,\,\,Z_2^0=\sqrt{3}(2\rho^2-1).\nonumber
\end{eqnarray}
Putting the ansatz Eq.~(\ref{ansatz0}) with
\begin{equation}
C=c_1 R_1^{1}+c_2 R_3^{1}\,,\,\,
S=s_1 R_1^{-1}+s_2 R_3^{-1}\,,\,\,
\tilde{T}=T_0 R_0^{0}+T_2 R_2^{0}\nonumber\\
\end{equation}
into the BC (\ref{apprBC}) and neglecting higher modes ($m\geq2$) 
yields the three conditions
$C(1)-T^{-}=0$, $S(1)=0$ and $\tilde{T}(1)-T^+=0$,
that can be solved for $c_2$, $s_2$ and $T_2$
to yield the following ansatz, fulfilling the BCs (up to $m=1$):
\begin{eqnarray}
T&=& s_1\left[ Z_1^{-1}-\frac{1}{\sqrt{2}}Z_3^{-1}\right] 
+c_1\left[Z_1^1 -\frac{1}{\sqrt{2}}Z_3^1\right]  +\frac{T^-}{2\sqrt{2}}  Z_3^1\nonumber\\
&&+T_0\left[Z_0^0 - \frac{1}{\sqrt{3}} Z_2^0\right]
+ \frac{1}{\sqrt{3}}T^+Z_2^0
\end{eqnarray}

We now insert this ansatz into Eq.~(\ref{rescaled}) and project on the different modes via
$\int_0^1 ... Z_1^{-1} \rho d\rho$ to project on $Z_1^{-1}$
(and analogously for $Z_1^{1}$ and $Z_0^0$). 
Using the relations for the Laplacian of the Zernike polynomials,
$\Delta Z_0^0=0,\Delta Z_1^1=0, \Delta Z_3^1=24Z_1^1 , \Delta Z_2^0=8 Z_0^0$,
as well as orthogonality, yields
\begin{eqnarray}
\dot{s}_1&=&-\frac{24D}{\sqrt{2}R^2} s_1 -\omega c_1\,,\nonumber\\
\dot{c}_1&=&-\frac{24D}{\sqrt{2}R^2}\left( c_1 -\frac{T^-}{2} \right)  +\omega s_1\,,\nonumber\\
\dot{T}_0&=&-\frac{8D}{\sqrt{3}R^2}\left( T_0-T^+\right) \,. 
\end{eqnarray}
Identifying $x=\alpha s_1$, $z=\alpha c_1$ and $\bar\psi=T_0$, this can be written as in the main text, i.e.
\begin{eqnarray}
\dot x&=&-rx -\omega z\,,\\
\dot z&=&-rz +\omega x +p\,,\\
\dot{\bar\psi}&=&-r_\psi\bar\psi+p_\psi\,,\label{storage}
\end{eqnarray}
with the rates 
\begin{equation}
r=\frac{24D}{\sqrt{2}R^2}\,\,,\,\,\,
p= \frac{24D}{\sqrt{2}R^2}\frac{\alpha(T_s-T_{air})}{4}
\end{equation}
for the $x,z$-modes (where $\alpha$ is the thermal expansion coefficient)
and
\begin{equation}
r_\psi=\frac{8D}{\sqrt{3}R^2}\,\,,\,\,\,
p_\psi=\frac{8D}{\sqrt{3}R^2}\frac{T_s+T_{air}}{2}\,, 
\end{equation}
for the storage mode.

As it should be, the stationary state for the storage mode is
$\bar\psi=\frac{p_\psi}{r_\psi}=\frac{T_s+T_{air}}{2}$,
i.e.~the mean temperature.
The stationary heat conducting state is given by
$z=\frac{p}{r}=\frac{\alpha(T_s-T_{air})}{4}$,
i.e.~proportional to the mean temperature difference.

\section*{Appendix C: Asymmetric thermal drive}

As discussed in the main text, a more realistic boundary condition is
\begin{equation}
 \frac{l(\phi)}{R}\frac{d}{d\rho}T(1,\phi)+\left[T(1,\phi)-T_{ext}(1,\phi)\right]=0.
\end{equation}
Here $l(\phi)$ is a thermal length scale describing convective transport in the fluid medium (air)
at the top, see Ref.~\cite{Baumann}. We assume $l(\phi=\pi)=l_{th}$ at the top and
at the bottom $l(\phi=0)=0$, i.e.~still perfect thermal contact at the metal surface.
The symmetry-adapted approximated version of this BC then reads
\begin{equation}
 l(1,\phi)=(1-\cos\phi)\frac{l_{th}}{2}
\end{equation}
and $T_{ext}(1,\phi)$ as defined in Eq.~(\ref{apprBC}).
Using the same Zernike-mode expansion and projection yields
the equations 
\begin{eqnarray}
\dot{x}&=& -r x -\omega z,\\
\label{fullz}
\dot{z}&=& -r z +p + \omega x + \chi\left(  -r_\psi\bar\psi +p_\psi  -\epsilon z   \right),\\
\dot{\bar{\psi}}&=&-r_\psi\bar\psi +p_\psi   -\epsilon z.
\end{eqnarray}
Now $r$, $p$, $r_\psi$, $p_\psi$ are all functions of $\lambda=\frac{l_{th}}{R}$ that can be given explicitly. 
More importantly, a coupling between $z$ and $\bar\psi$ arises, with
\begin{eqnarray}
 \epsilon=\frac{8}{\sqrt{3}}\frac{D}{\alpha R^2}\frac{\frac{6\lambda}{7\lambda+2}}
{1+\frac{3}{4}\lambda\frac{7\lambda+4}{7\lambda+2}}\,\,\,,\,\,\,
\chi=\alpha\frac{9\sqrt{3}}{2\sqrt{2}}\frac{\lambda}{7\lambda+2}.
\end{eqnarray}
It can be seen that for $l,\lambda\rightarrow0$ we regain the symmetric case
since $\epsilon,\chi\rightarrow0$. The term in brackets in Eq.~(\ref{fullz})
can be interpreted as a temporal modulation of the driving term $p$ of the $z$-mode,
as discussed in the main text (where the contribution $\epsilon z$ was omitted to simplify the 
discussion).

\section*{Conflicts of interest}

There are no conflicts to declare.

\section*{Acknowledgements}
We thank Vincent Le Houerou and Fabien Montel for fruitful discussions. This article was created entirely with Free and Open-Source Software (FOSS): ImageJ/FiJi, Jupyter-Lab, QtiPlot (0.9.8),  Maxima-CAS/wxMaxima, Python/Matplotlib, GNU-Octave, LibreOffice Draw, Inkscape, OpenShot, Kdenlive, Lyx and TexStudio. The authors acknowledge the effort of numerous FOSS programmers without whom our work would be very difficult and much less enjoyable.



\balance


\bibliographystyle{rsc} 

\end{document}